\documentclass{amsart}
\usepackage{amsmath,amssymb,latexsym, amsmath,graphics}

\advance \topmargin by -\headheight
\advance \topmargin by -\headsep
\evensidemargin \oddsidemargin
\newtheorem{thm}{Theorem}[section]

\newtheorem{prop}[thm]{Proposition}

\def\la{\lambda}

\DeclareMathOperator{\Jac}{Jac}

\DeclareMathOperator{\Res}{Res}

\begin{document}
\title[Dynamics on strata of trigonal Jacobians ]
{Dynamics on strata of trigonal Jacobians and some integrable problems of rigid body motion}

\author{H.W. Braden}
\address{School of Mathematics, Edinburgh University, Edinburgh.}
\email{hwb@ed.ac.uk}
\author{V.Z. Enolski}
\address{School of Mathematics, Edinburgh University, Edinburgh.
 On the leave from Institute of Magnetism, National Academy of Sciences of
Ukraine, Kiev} \email{venolski@googlemail.com}
\author{Yu.N.Fedorov}
\address{Department of Mathematics I, Politechnic university of Catalonia, Barcelona}
 \email{Yuri.Fedorov@upc.es}

\begin{abstract}
We present an algebraic geometrical and analytical description of
the Goryachev case of rigid body motion. It belongs to a family of systems sharing the same properties:
although completely integrable, they are
not algebraically integrable, their solution is not meromorphic in the complex time
and involves dynamics on the strata of the Jacobian varieties of trigonal curves.

Although the strata of hyperelliptic Jacobians have already appeared in the literature in the context of
some dynamical systems,
the Goryachev case is the first example of an integrable system whose solution involves a more general curve. Several
new features (and formulae) are encountered in the solution given in terms of sigma-functions of such a curve.
\end{abstract}

\maketitle

\section{Introduction}
Most of the known finite-dimensional integrable systems of classical mechanics and mathematical physics
are also algebraically completely integrable: following \cite{AvM1}, their invariant tori can be extended to specific
complex tori, Abelian varieties, and the complexified flow is a straight-line flow on them.
As a direct consequence of this property, all the solutions of such systems are meromorphic functions of the
complex time, and can be described explicitly in terms of theta-functions or generalized theta-functions. The property of meromorphicity led to the Kovalevskaya--Painlev\'e integrability test, which was effectively applied to detect several integrable cases. In some cases, such as the famous Neumann system describing the motion of a point on a sphere with a quadratic potential, or the Steklov--Lyapunov integrable case of the Kirchhoff
equations, the complex tori are Jacobians of hyperelliptic curves (or possibly coverings of the Jacobians).
In other, more complicated situations, for example the Frahm--Manakov top on $so(n)$ or the Kovalevskaya top, the complexified tori are not Jacobians but  Abelian subvarieties thereof, with a non-principal polarization (Prym varieties)\footnote{
These can be related to other Jacobians in the case of dimension 2.}.

On the other hand, there are many other systems, including generalizations of the above ones, which preserve integrability but lose the meromorphicity of the complex solutions. In such systems the genus of the underlying algebraic curve (often the spectral curve) is greater than the dimension of the invariant tori, and the latter are
certain non-Abelian subvarieties (strata) of Jacobians. The algebro-geometric properties of such systems and the nature of the singularities of their complex solutions have been described in \cite{Vanh, abendfed00, EPR03, fg07, eekl93,eekt94}. Until recently, however, the only examples known were of systems related to the strata
of hyperelliptic Jacobians.

In present paper we consider the first example of a mechanical system whose complex invariant varieties are strata
of Jacobians of a non-hyperelliptic curve, here a trigonal curve of genus 3 given by the equation \\ $y^3+p(x)y+q(x)=0$.
The latter appear in the reduction to quadratures of the integrable Goryachev case of the Kirchhoff equations \cite{Gor1}:
the quadratures involve 2 points on the genus 3 curve (which has no additional involution in general, so
not allowing reduction to a Prym) and so the quadratures lead to an incomplete Abel
map which cannot be inverted in terms of meromorphic functions. This means that, as in other
non-algebraic integrable systems, the Goryachev case cannot be detected by the Kovalevskaya--Painlev\'e test.
We emphasize that this example is not unique:
it is in fact a member of a family of integrable Hamiltonian flows on the sphere $S^2$ that have
extra cubic integrals and that were described recently in \cite{Val}, see also \cite{Ver_Tsi_12, Yeh02}.
In particular this family also includes the non-trivial case found by Dullin and Matveev in \cite{DM}.
As was shown in \cite{Tsi_05, Ver_Tsi}, most of the systems of the family are reduced to ``trigonal''
quadratures similar to those of the Goryachev system\footnote{There are several exceptions in the family: one of them
is the classical Goryachev--Chaplygin system \cite{Chapl,Gor2}, which is linearized on Jacobians of genus 2 hyperelliptic
curve.}. For concreteness, the present paper considers only the Goryachev system.

Our solution builds on the explicit description of Abelian functions of trigonal curves
and (in the terminology of \cite{bel99}) the more general $(n,s)$-curves. There has been a
resurgence of interest in this area with many new analytic results obtained
\cite{bel00, bl05,eemop07,ee09,nak10,eeo11} including the inversion of the Abel-Jacobi map on strata of Jacobians
\cite{bg06,matprev08,matprev11,nakyo12}. Although we will need to extend this work in various ways these new
studies are foundational to providing explicit solutions for non-algebraically completely integrable systems.

The paper is organized as follows.
In Section 2 we reproduce the reduction to quadratures of the Goryachev system first made in \cite{Ver_Tsi}
and interpret them as sums of two holomorphic differentials on a genus 3 trigonal curve $\mathcal{C}$,
also indicating its canonical form.
The original variables of the system are then expressed in terms of coordinates of two points on the curve.
In Section 3, following \cite{eemop07}, we describe in detail the inversion of the complete Abel map
on the (3,4)-trigonal curve. Here the main tools are the sigma-function of the curve and its
logarithmic derivatives (Kleinian-Weierstrass functions),
which are direct generalizations of the Weierstrass elliptic $\sigma$- and $\wp$-functions respectively
(see \cite{bel97, ba97}). We give a method of an effective calculation of the vector of Riemann constants for the trigonal Jacobians which allows us to calculate the sigma function explicitly by relating it to the corresponding theta-function.
We also present an analytic description of the Wirtinger strata in the Jacobian of $\mathcal{C}$ as zeros of
the sigma-function and of some of its derivatives.
Section 4 describes the inversion of the incomplete Abel map.  The formal explicit solution to the inversion problem
is obtained from the formulae of the previous section (inversion of the complete map) by a certain limiting procedure.
The resulting solution is given in terms of the sigma-function and its derivatives whose arguments are
restricted to the 2-dimensional stratum in the Jacobian of $\mathcal{C}$.
(Note that for the case of a cyclic trigonal curve of genus 4 similar results were obtained in \cite{bg06}.)

Then global analytical properties of the solutions as functions of the complex time are described.
We show that  these functions have an infinite number of branch points and are single-valued only
on an infinite ramified covering of the complex time plane.
Finally the local singularities of the complex solutions are described
in Section 5 by  using the expansion of the sigma-function near generic and special points of the
curve $\mathcal{C}$. An Appendix contains some rather long and technical proofs of two theorems.

\section{The Goryachev integrable case. Separation of variables and reduction to quadratures.} Recall that
the classical Kirchhoff equations describing the motion of a rigid body in an ideal fluid have (in an
appropriate coordinate frame) the form
\begin{equation} \label{K}
\begin{aligned}
 \dot J & = J\times \frac{\partial H}{\partial J} + \gamma \times \frac{\partial H}{\partial \gamma} , \\
 \dot \gamma & = \gamma \times \frac{\partial H}{\partial J} ,
\end{aligned}
\end{equation}
where $J=(J_1, J_2, J_3)^T$  and
$\gamma=(\gamma_1,\gamma_2,\gamma_3)^T$  are the angular and linear momentum respectively and $H(J,\gamma)$ is the Hamiltonian, which is also
a first integral. In addition to the Hamiltonian the equations always possess the integrals (Casimir functions)
$$
C_1= \langle J,\gamma\rangle, \quad C_2= \langle \gamma, \gamma\rangle .
$$

Apart from the well known integrable cases of Kirchhoff, Clebsch, Steklov and Lyapunov (and their gyroscopic generalizations)
there are some further special cases of integrability cases where an additional integral exists only under the condition $C_1=0$.
In the most classical case found by D. Goryachev \cite{Gor1} and S. Chaplygin \cite{Chapl} the extra integral is cubic in
$J$. In this case Chaplygin himself \cite{Chapl} gave a separation of variables and reduced the system to quadratures
containing integrals on a hyperelliptic genus 2 curve. A detailed algebro-geometric description of the complex
invariant manifolds was made in \cite{BvM}. Further rather exotic special cases of integrability also exist
for which neither separation of variables nor explicit solution were known until recently.
Here we concentrate on the Goryachev case \cite{Gor2} which was reduced to quadratures
in \cite{Ver_Tsi} by using a bi-Hamiltonian structure and the corresponding separating Darboux--Nijenhuis variables.

The Goryachev case, the focus of this paper, has Hamiltonian $H_1=H$ and extra integral $H_2$ that take the form
\begin{align}\begin{split}
H_1 & = J_1^2+J_2^2+ \frac 4 3 J_3^2 + \frac {a\gamma_1+b}{\gamma_3^{2/3}}  , \quad \text{ $a,b$ being arbitrary constants},  \\
H_2 & =  -\frac 23 J_3 (J_1^2+J_2^2+ \frac 8 9 J_3^2 +  \frac {a \gamma_1+ b}{\gamma_3^{2/3}} )+ a \gamma^{1/3}_3 J_1 .\end{split}
\label{ints}
\end{align}
The corresponding Kirchhoff equations are then
\begin{align}\begin{split}
\dot J_1 & = \frac 23 J_2 J_3 - \frac 23 \gamma_2 (a \gamma_1+b) \gamma_3^{-5/3} ,  \\
\dot J_2 & = - \frac 23 J_1 J_3 + a \gamma_3^{1/3} + \frac 23 \gamma_1 (a \gamma_1+b) \gamma_3^{-5/3} ,  \\
\dot J_3 & = - a \frac {\gamma_2}{\gamma_3^{2/3}},  \\
\dot \gamma_1 & = 2 (\frac 43 J_3 \gamma_2 - J_2 \gamma_3),  \\
\dot \gamma_2 & = 2 (-\frac 43 J_3 \gamma_1 + J_1 \gamma_3),  \\
\dot \gamma_3 & = 2 (\gamma_1 J_2- \gamma_2 J_1).  \end{split}\label{eq_Gor}
\end{align}

Without loss of generality one can set $C_2=\langle \gamma, \gamma\rangle =1$.
Then, since $C_1=0$, equations \eqref{K} can be
reduced to a Hamiltonian system on the cotangent bundle of the unit sphere
$S^2=\{\langle \gamma, \gamma\rangle =1\}$ with coordinates and momenta
\begin{equation} \label{S2_var}
u=\gamma_3, \quad p_u=\frac{J_1 \gamma_2 - J_2 \gamma_1}{\gamma_1^2+\gamma_2^2} , \quad
\phi=\arctan (\gamma_1/\gamma_2), \quad  p_\phi=-J_3.
\end{equation}
In terms of these the original variables become
\begin{gather}
J_1 =\dfrac{u \sin\phi\,p_\phi+\cos\phi(1-u^2)\,p_u}{\sqrt{1-u^2}}\,,\qquad
J_2 =\dfrac{ u\cos\phi\,p_\phi-\sin\phi(1-u^2)\,p_u}{\sqrt{1-u^2}}\,, \notag \\
 \gamma_1 = \sqrt{1-u^2}\sin\phi,\qquad  \gamma_2 = \sqrt{1-u^2}\cos\phi,\qquad  \gamma_3 = u\,. \label{**}
\end{gather}
The paper \cite{Ver_Tsi} introduced separating variables $q_1$, $q_2$ as the roots of the polynomial
$$
A(\la) = \la^2+ u^{1/3} \left( \frac{u p_\phi}{1-u^2} - \imath p_u \right) \la
- \frac{\imath a e^{\imath \phi}}{\sqrt{1-u^2}} , \qquad \imath=\sqrt{-1}.
$$
Observe this polynomial depends not only on the coordinates $u, \phi$ of $S^2$ but also on their momenta $p_u$, $p_\phi$.
Following  \cite{Ver_Tsi}, in the Darboux coordinates $q_i, p_i$ such that
\begin{equation} \label{conj}
\{ q_i, p_j\} = \delta_{ij}, \quad  \{ q_i, q_j\} =  \{ p_i, p_j\} =0,
\end{equation}
one has
\begin{gather} \label{rel}\begin{split}
 p_\phi = \imath\, q_1 q_2 \frac{ p_2-p_1  }{q_1-q_2}, \quad
u = \left[ -\frac 23 \imath \frac{q_1p_1-q_2p_2 }{q_1-q_2}  \right]^{3/2} , \\
u^{1/3} \left( \frac{u p_\phi}{1-u^2} - \imath p_u \right) = -q_1-q_2 , \quad
\frac{\imath a e^{\imath \phi}}{\sqrt{1-u^2}} = -q_1 q_2 .
\end{split}
\end{gather}
Here one should stress that in the real case the coordinates $q_1,p_1, q_2,p_2$ are complex. In the sequel, by default,
we consider all the variables as complex, leaving the analysis of real conditions to a separate study in the future.
In particular, $q_i, p_i$ will be regarded as Darboux coordinates on the complexified cotangent bundle
$T^* \, S^2$.

Under the above substitution the two Hamiltonians take the St\"ackel form
\begin{gather}
\begin{split}
H_1 &= (S^{-1})_{11} U(p_1,q_1)+  (S^{-1})_{21} U (p_2,q_2) , \\ 
H_2 &= (S^{-1})_{12} U(p_1,q_1)+  (S^{-1})_{22} U (p_2,q_2) , \\ 
U(q,p)&= \frac{3}{2 \imath q^2 p} \left(-\frac {8}{27}\imath p^3 q^4 + \la^4+\frac14a^2- b \la^2 \right),
\end{split}
 \label{hams}
\end{gather}
with the St\"ackel matrix
$$
S=\begin{pmatrix}  1 & 1 \\ 3\imath/(2p_1 q_1) & 3\imath/(2p_2 q_2) \end{pmatrix}.
$$
Setting for convenience $\la_i=q_i$, $\mu_i = 2/3\, \imath q_i p_i$, we get
$$
U(q,p)=\bar U(\la,\mu)= \frac{1}{\la \mu} (\la \mu^3 + \la^4+a^2/4- b \la^2), \quad
S=\begin{pmatrix}  1 & 1 \\ -1/\mu_1 & -1/\mu_2 \end{pmatrix}
$$
and then observe that the above relations are equivalent to
$$
H_1- H_2 \frac 1 {\mu_1}= \bar U(\la_1,\mu_1) , \quad H_1- H_2 \frac 1 {\mu_2}= \bar U(\la_2,\mu_2)
$$
Then, upon fixing the values of the integrals, $H_1=h_1, H_2=h_2$,
the pairs $(\la_i,\mu_i)$ are subject to the algebraic relation
$$\mu h_1-h_2 = \mu^3+ \la^3 + \frac{a^2}{4\la}- b \la$$
or
\begin{equation} \label{curve}
F= \la^4- b\la^2 + (\mu^3-h_1 \mu+h_2)\la + a^2/4 =0 .
\end{equation}

As was also mentioned in \cite{Ver_Tsi}, equation \eqref{curve}
defines an algebraic curve $\mathcal{C}'\subset {\mathbb C}^2(\la,\mu)$, which, for generic values of $h_1, h_2$ is smooth, has genus 3, and is non-hyperelliptic, i.e.,
cannot be transformed to the form $y^2=P(x)$ by a birational change of coordinates.
A basis of holomorphic differentials on $\mathcal{C}'$ is
\begin{equation} \label{difs}
\omega_1 = \frac{d\la }{ \partial F/\partial\mu }, \quad \omega_2 =\frac{\la d\la }{\partial F/\partial\mu},
 \quad \omega_3 =\frac{\mu \, d\la }{ \partial F/\partial\mu }, \qquad \frac{\partial F}{\partial\mu}= \la(3\mu^2-h_1).
\end{equation}

\subsection{The quadratures.} Let $t_1, t_2$ denote the time of the flows on $T^* S^2$ defined respectively by
the Hamiltonians $H_1$ and $H_2$.
To describe the evolution of $q_i=\la_i$ with $H_2$,
we use the bracket \eqref{conj} and the expressions \eqref{hams} to obtain
\begin{align*}
\frac{d}{d \,t_2} \la_1 & = \frac{\partial H_2}{ \partial p_1}=
\frac{\partial H_2}{ \partial \mu_1} \frac 23 \imath \la_1 =
 \left[ \left( \frac{\mu_2}{\mu_1-\mu_2} -  \frac{\mu_1\mu_2}{(\mu_1-\mu_2)^2}  \right) ( \bar U_1-\bar U_2 ) +
\frac{\mu_1\mu_2}{\mu_1-\mu_2}  \frac{\partial \bar U_1}{\partial \mu_1}  \right] \frac 23 \imath \la_1 \\
 & =   \frac 23 \imath \la_1 \frac{\mu_2}{\mu_1-\mu_2}
\left ( \bar U_1 + \mu_1 \frac{\partial \bar U_1}{\partial \mu_1} -
\underbrace{\frac1{\mu_1-\mu_2}(\mu_1 \bar U_1-\mu_2 \bar U_2)}_{H_1(\la,\mu)}  \right), \qquad
\bar U_i=\bar U(\la_i,\mu_i).
\end{align*}
In view of the expression for $F$ in \eqref{curve} this is equivalent to
\begin{equation} \label{der_la_2}
 \frac{d\la_1 }{d \,t_2} = \frac 23 \imath   \frac{\mu_2}{\mu_1-\mu_2}\frac{\partial F(\la_1,\mu_1)}{\partial\mu_1}, \quad \text{and, similarly,} \quad
 \frac{d\la_2 }{d \,t_2} = \frac 23 \imath   \frac{\mu_1}{\mu_2-\mu_1}\frac{\partial F(\la_2, \mu_2)}{\partial\mu_2} .
\end{equation}
Similarly one obtains for the flow with the quadratic Hamiltonian $H_1$
\begin{equation} \label{der_la}
 \frac{d\la_1}{d \,t_1} = -\frac 23 \imath \frac{1}{\mu_1-\mu_2} \frac{\partial F(\la_1,\mu_1)}{\partial\mu_1}, \quad
 \frac{d\la_2}{d \,t_1} = -\frac 23 \imath  \frac{1}{\mu_2-\mu_1} \frac{\partial F(\la_2, \mu_2)}{\partial\mu_2} ,
\end{equation}
and also
\begin{equation}
 \frac{d\mu_1  }{d \,t_1} =  \frac{\partial \mu_1} {\partial\lambda_1 } \frac{d\la_1}{d \,t_1}
= \frac 23 \imath \frac{1}{\mu_1-\mu_2} \frac{\partial F(\la_1,\mu_1)}{\partial\lambda_1} , \quad
   \frac{d\mu_2}{d \,t_1} = \frac 23 \imath \frac{1}{\mu_2-\mu_1} \frac{\partial F(\la_2,\mu_2)}{\partial\lambda_2}.
\label{der_mu}
\end{equation}
Expressions \eqref{der_la}, \eqref{der_la_2} yield the following quadratures in the differential form
\begin{equation} \label{quads}
\begin{aligned}
  \frac {d\la_1} {\partial F(\la_1,\mu_1) /\partial\mu_1 }  + \frac {d\la_2}{\partial F(\la_2,\mu_2) /\partial\mu_2 } &
= - \frac 23 \imath  \, dt_2 \\
  \frac {\mu_1 d\la_2}{\partial F(\la_1,\mu_1) /\partial\mu_1}+\frac {\mu_2 d\la_2}{\partial F(\la_2,\mu_2) /\partial\mu_2 } &
= -\frac 23 \imath  \, dt_1 .
\end{aligned}
\end{equation}
We will return to these later.

\subsection{The original variables in terms of the separating ones.}
From \eqref{rel} and other formulae from the paper \cite{Ver_Tsi} one has
\begin{align}
\begin{split}
\gamma_3^2 & = - \left( \frac{\mu_1-\mu_2}{\la_1-\la_2} \right)^3 , \quad
\gamma_3^{2/3} = - \frac{\mu_1-\mu_2}{\la_1-\la_2} ,  \\
J_3 & =- p_\phi = - \frac 32 \frac {\la_1\mu_2-\la_2 \mu_1}{\la_1-\la_2},  \\
\exp(\imath \phi) & = \frac{2\imath}{a} \, \la_1\la_2 \sqrt{1+  \left( \frac{\mu_1-\mu_2}{\la_1-\la_2}\right)^3  },
\\
\gamma_2+ \imath \gamma_1 & = \frac{2\imath}{a} \, \la_1\la_2 \left(1+ \left( \frac{\mu_1-\mu_2}{\la_1-\la_2}\right)^3 \right), \quad
 \gamma_2 -\imath \gamma_1 = \frac{a}{2\imath \, \la_1\la_2} ,  \\
J_1 + \imath J_2 & = -\imath \left( \imath p_u - \frac {u}{1-u^2} p_\phi \right) e^{-\imath \phi} \sqrt{1-u^2}  \\
& =-\frac a 2 \,\frac{\la_1+\la_2}{\la_1 \la_2} \left(- \frac{\mu_1-\mu_2}{\la_1-\la_2} \right)^{-1/2}\, , \\
 J_1 - \imath J_2 & = \frac{1}{\gamma_2-\imath \gamma_1}
\left((J_1+\imath J_2) (\gamma_2+\imath\gamma_1)+ 2 \imath J_3\gamma_3 \right) \quad \text{(due to the condition
$J\cdot \gamma=0$)}
 \\
& ={\displaystyle \frac {2\, \lambda_1 \, \lambda_{2}\, \left(
 \!  \left(  \! 1 + {\displaystyle \frac {({\mu_{1}} - {\mu_{2}})^{3}
}{({\lambda_{1}} - {\lambda_{2}})^{3}}}  \!  \right) \,({\lambda_{1}} + {\lambda_{2}}) +
{\displaystyle \frac {3\,({\mu_{1}} - {\mu_{2}})^{2}\,({\lambda_{1}}\,{\mu_{2}}
- {\mu_{1}}\,{\lambda_{2}})}{({\lambda_{1}} - {\lambda_{2}})^{3}}}  \!  \right) }{
a\,\sqrt{ - {\frac {\mu_1 - \mu_2}{\lambda_1 - x_2 }} }}} \end{split}\label{exp_lambdas}
\end{align}
Substituting the above expressions into the integrals \eqref{ints} and using the equation of the curve
\eqref{curve} for each pair $(\la_i,\mu_i)$  one may verify the identities $H_1=h_1, H_2=h_2$. Thus to
solve for the original system \eqref{K}, \eqref{ints} it suffices to solve for the pairs $(\la_i,\mu_i)$.

\subsection{Canonical form of the curve and of the quadratures.}
Rather than using the variables $(\la_i,\mu_i)$ directly we now make one final birational transformation
to bring the curve $\mathcal{C}'$ to a canonical form. This allows us to make connection with the literature
on $(n,s)$ curves and so permits us to solve for the motion in terms of the multi-dimensional $\sigma$-function.

By making the birational change
\begin{equation}
\lambda= \frac{1}{x}\sqrt{\frac{a}{2}}, \quad \mu=-\frac{y}{x}\sqrt{\frac{a}{2}} \label{subs}
\end{equation}
the curve \eqref{curve} can be transformed to the {\it canonical trigonal form}
with respect to $y$
\begin{equation} \label{curve1}
G (x,y)= y^3 - 2 \frac{h_1}{a} x^2\, y - \left( x^4+ \frac{2 \sqrt{2} h_2}{a^{3/2}}  x^3
- 2 \frac{b}{a} x^2 +1 \right) =0 .
\end{equation}
We will refer to this curve as $\mathcal{C}$. In the terminology of
\cite{bel00} (see also the next section) this is a (3,4)-curve having
one infinite branch point $\infty$, where all the 3 sheets of the covering
$\mathcal{C}\to {\mathbb P}\sp1=\{ x\}$ come together.
Under the above transformation the holomorphic differentials $\omega_1, \omega_2, \omega_3$ in \eqref{difs}
take the following respective forms,
\begin{equation} \label{difs_t}
\Omega_2 = - \frac{x\, \mathrm{d} x }{\partial G/\partial y}, \quad
\Omega_1 = \frac{\mathrm{d}x }{\partial G/\partial y},
 \quad \Omega_3 = \frac{y \, \mathrm{d}x}{\partial G/\partial y }, \qquad
\frac{\partial G}{\partial y}= 3y^2-\frac {2h_1}{a} x^2.
\end{equation}
Then, in the new coordinates, the quadratures \eqref{quads} take the form
\begin{equation} \label{quads2}
\begin{aligned}
  \frac {x_1\, \mathrm{d} x_1} {\partial G(x_1,y_1) /\partial y_1}
+ \frac {x_2 \mathrm{d} x_2}{\partial G(x_2,y_2)/\partial y_2 } & = -\frac 23 \imath  \, \mathrm{d}t_2 ,\\
  \frac {y_1 \mathrm{d}x_1}{\partial G(x_1,y_1) /\partial y_1 }
 + \frac {y_2 d x_2}{\partial G(x_2,y_2) /\partial y_2 } & = -\frac 23 \imath  \, \mathrm{d}t_1 .
\end{aligned}
\end{equation}
We again stress that, although in the real case the curve \eqref{curve1} is real, the new separating variables $x_1, x_2$
are complex. The description of their behaviour in the real case could be an object of a separate study.

\subsection{Non-algebraic integrability.} We are now in the situation described
in the introduction. The genus of the curve $\mathcal{C}$ is greater than that of the dimension of generic invariant tori of the system: here we have 2 separating variables while $\mathcal{C}$ is of genus 3.
Such a situation occurs in many algebraically integrable systems, for example the Clebsch integrable case of the Kirchhoff equations or the Kovalevskaya top. In these examples however, although the genus of the underlying curve (the spectral curve of the corresponding Lax representation) is greater than the number of
degrees of freedom, the relevant curves possess an additional involution which extends to the Jacobian variety.
Then, as a rule, the complex invariant manifolds of the systems turn out to be 2-dimensional
Abelian (Prym) subvarieties of the Jacobians, whose real part gives the invariant tori.
This however is not the case  for the Goryachev system on $T^*S^2$ we are considering. For generic $h_1, h_2$
the curve $\mathcal{C}$ has no further symmetries, and the differentials in \eqref{quads2} do not reduce to those of a genus 2 curve.
As we shall see below, this pathological property means such systems are not algebraically completely integrable.
In particular, their complex invariant manifolds are non-Abelian subvarieties of the Jacobian of $\mathcal{C}$ and the variables
$J_i, \gamma_i$ are not meromorphic functions of the complex times $t_1, t_2$.

In terms of the differentials \eqref{difs_t} the quadratures \eqref{quads2} may be extended and
written in the following integral form
\begin{equation} \label{AM2}
\int_{\infty}^{P_1} (\Omega_1, \Omega_2,\Omega_3)^T +  \int_{\infty}^{P_2} (\Omega_1, \Omega_2,\Omega_3)^T
= \left\{
\begin{aligned} u_1 &  \\ u_2 &=-2\imath t_2/3+u_{20} & \\ u_3& = - 2\imath  t_1/3 +u_{30} \end{aligned} \right. , \quad
P_i=(x_i, y_i)\in \mathcal{C} \, ,
\end{equation}
where $u_{20}, u_{30}$ are constant phases and the coordinate $u_1$ is a transcendental function of $u_2, u_3$,
whose properties will be described in the next sections.
Thus \eqref{AM2}
defines a map of the symmetric product $\mathcal{C}\times \mathcal{C}$ to a codimension one subvariety (stratum)
of the Jacobian variety of $\mathcal{C}$. To invert the map, i.e., to express symmetric
functions of the coordinates $(x_1, y_1), (x_2, y_2)$ and consequently the variables $J_i$, $\gamma_i$ in terms of $u_2, u_3$,
at least locally, we next recall some basic facts about the standard Jacobi inversion problem associated to
trigonal curves.

\section{Jacobi inversion problem for the genus three trigonal curve}
The curve \eqref{curve1} belongs to a class of $(n,s)$-curves.
These are smooth curves with $s>n\ge2$ and  $\mathrm{gcd}(n,s)=1$ that have one point at infinity and whose affine part may be
defined by an equation
$$
y^n-x^s-\sum_{\alpha,\beta} \nu_{\alpha n+\beta s} x^\alpha y^\beta =0, \qquad 0\le \alpha <s-1, \quad 0\le \beta <n-1.
$$
The curves are of genus $g=(n-1)(s-1)/2$ and their properties and relation to integrable hierarchies of KP type are widely discussed in the literature.

Below we concentrate on the trigonal (3,4)-curve $\widetilde{\mathcal{C}}$, which we write in the canonical form
\begin{equation}
f(x,y)=y^3+(\mu_2x^2+\mu_5x+\mu_8)y-(x^4 +\mu_3 x^3 +\mu_6 x^2 +\mu_9 x +\mu_{12})=0 ,
\label{trig}
\end{equation}
where $\mu_j$ are parameters. This is more general than our curves \eqref{curve1} and we will specialize in due course.
The curve has one infinite point $\infty$, where all 3 sheets of the covering ${\mathcal C} \to {\mathbb P}=\{x\}$
come together.
Let $\xi=x^{-1/3}$ be a local coordinate in a neighborhood of $\infty$. That is, $\xi(\infty)=0$, and the coordinates in the vicinity of this point have expansions
\begin{equation} \label{exp_y}
x = \frac{1}{\xi^3}, \quad y = \frac{1}{\xi^4}- \frac{\mu_2}{3} \frac{1}{\xi^2}+ \frac{\mu_3}{3} \frac{1}{\xi} + O(\xi).
\end{equation}
Choose the vector of holomorphic differentials $\boldsymbol{\Omega}=(\Omega_1,\Omega_2,\Omega_3)^T$,
\begin{equation} \label{OMS}
 \boldsymbol{\Omega} = \frac{d x}{f_y(x,y)} \left(\begin{array}{c} 1 \\ x \\y  \end{array}\right) \, .
\end{equation}
Near $\infty$, they admit the expansions
\begin{align}
\Omega_1 &=-\xi^4 \left(1+ \frac{\mu_2}{3} \xi^2 - \frac{2 \mu_3}{3} \xi^3 + O(\xi^5))\right) d\xi, \notag \\
\Omega_2 & = -\xi \left(1+ \frac {\mu_2}{3}\xi^2 - \frac {2 \mu_3}{3}\xi^3 + O(\xi^5)\right )  \mathrm{d}\xi,
\label{expan_difs} \\
\Omega_3 & = -\left(1 -\frac{\mu_3}{3} \xi^3 - \frac{\mu_2^2}{9} \xi^4 + O(\xi^6)\right ) \mathrm{d}\xi, \notag
\end{align}
so the orders of their zeros at $\infty$ decrease.

Next, choose a canonical basis of cycles of $H_1(\mathcal{C},\mathbb{Z})$
\[
(\mathfrak{a},\mathfrak{b})=( \mathfrak{a}_1,\mathfrak{a}_2,\mathfrak{a}_3;
\mathfrak{b}_1,\mathfrak{b}_2,\mathfrak{b}_3  ), \quad \text{such that} \quad
\mathfrak{a}_i\circ \mathfrak{a_j} = \mathfrak{b}_i\circ \mathfrak{b_j}=0,
\quad \mathfrak{a}_i\circ \mathfrak{b_j}= - \mathfrak{b}_i\circ \mathfrak{a_j}=1
\]
and introduce matrices of periods of the above differentials
\begin{equation} \label{periods_AB}
\mathcal{A}=\left(  \oint_{\mathfrak{a}_j}  \Omega_i \right)_{i,j=1,2,3},\quad
\mathcal{B}=\left(  \oint_{\mathfrak{b}_j}  \Omega_i \right)_{i,j=1,2,3}\, .
\end{equation}

Throughout the whole paper we will use two normalization of periods: the first, given above,
$(\mathcal{A},\mathcal{B})$, where we have specified the differentials; the second utilizes the so called
 $\mathfrak{a}$-normalized differentials for which the matrix of periods takes the form $(1_3,\tau)$.
Here $\tau$ is the Riemann period matrix
\begin{equation}
\tau=\mathcal{A}^{-1}\mathcal{B}, \quad  \tau^T=\tau, \quad \mathrm{Im}\, \tau >0 \, . \label{tau}
\end{equation}
We denote the $\mathfrak{a}$-normalized holomorphic differentials by
\begin{equation}
\overline{\boldsymbol{\Omega}}= \mathcal{A}^{-1}\boldsymbol{\Omega}\quad
\Leftrightarrow\quad  \oint_{\mathfrak{a}_j} \bar\Omega_i=\delta_{ij}, \quad i,j=1,2,3.
\end{equation}
The first normalization is used in the definition of the $\sigma$-function while the second
in the definition of Riemann's theta-function $\theta(\boldsymbol{z};\tau)$ associated with $\mathcal{C}$ and the
Riemann period matrix $\tau$ given in (\ref{tau}); it is defined by the series
\begin{gather}
\theta(\boldsymbol{v};\tau) =\sum_{\boldsymbol{n}\in \mathbb{Z}^3} \mathrm{exp}
\left\{ \imath \pi  \boldsymbol{n}^T \cdot\tau  \cdot\boldsymbol{n}
+ 2 \imath \pi  \boldsymbol{v}^T  \cdot\boldsymbol{n} \right\} \, .
\end{gather}

With these normalizations we define the Jacobi variety $\mathrm{Jac}(\mathcal{C})$ of the curve $\mathcal{C}$
to be the quotient $\mathrm{Jac}(\mathcal{C}) =\mathbb{C}^3(u_1, u_2, u_3)/\{ \mathcal{A}\oplus \mathcal{B}\}$
and also $\overline{\mathrm{Jac}}(\mathcal{C}) =\mathbb{C}^3(v_1, v_2, v_3)/\{1_3\oplus \tau\}$,
where the vectors $\boldsymbol{u}$ and $\boldsymbol{v}$ are related by $\boldsymbol{v}=(v_1, v_2, v_3)^T=\mathcal{A}^{-1}\boldsymbol{u}$.

For a positive divisor $\mathcal{D}= P_1+ P_2+P_3$, $P_i=(x_i,y_i) \in \mathcal{C}$, we
consider {\it the Abel map} with a base point $P_0$,
$\boldsymbol{\mathfrak{A}}: \mathcal{C} \times \mathcal{C} \times \mathcal{C} \longrightarrow \mathrm{Jac}(\mathcal{C})$,
given by
\begin{equation}
\boldsymbol{\mathfrak{A}}:\mathcal{D} \longrightarrow
\int_{P_0}^{P_1} \boldsymbol{\Omega}+ \int_{P_0}^{P_2} \boldsymbol{\Omega}+ \int_{P_0}^{P_3}\boldsymbol{\Omega}
=\boldsymbol{u}, \quad \text{or, equivalently,} \quad
\int_{P_0}^{P_1} \overline{\boldsymbol{\Omega}}+ \int_{P_0}^{P_2} \overline{\boldsymbol{\Omega}}+
\int_{P_0}^{P_3} \overline{\boldsymbol{\Omega}} =\boldsymbol{v},   \label{JIP}
\end{equation}
Henceforth we assume that $P_0=(\infty,\infty)$.
The Jacobi inversion problem refers to the inversion of Abel map. For the $g=3$ case being considered this is solved
in terms of Riemann's $\theta$-functions as follows.

\begin{thm} \textup{(Riemann)}
Let $\boldsymbol{e}$ be a vector such that the function $F(P)=\theta(\int_{\infty}^{P}\overline{\boldsymbol{\Omega}}-\boldsymbol{e};\tau)$ does not vanish identically. Then $F(P)$ has
exactly 3 zeros on $\mathcal{C}$, $P_1,P_2,P_3$, and these provide a solution of the Jacobi inversion problem
\begin{equation}
\int_{3\infty}^{{P}_1+{P}_2+{P}_3}{ \overline{\boldsymbol{\Omega}}}
= \boldsymbol{e}-\overline{\boldsymbol{K}}_{\infty},
\label{JIP'}
\end{equation}
where $\overline{\boldsymbol{K}}_{\infty}=( \overline{K}_1,\overline{K}_1,\overline{K}_3 )$
is the vector of Riemann constants with base point $\infty$ and whose components are given by
\begin{equation} \label{Riemann_K}
\overline{K}_{r}=\frac 12 (2\pi \imath+\tau_{rr})-{1\over 2\pi \imath}
\sum^{3}_{l\neq r} \left(\oint_{{\mathfrak a}_l} \overline{\Omega}_{l}(P)
\int^{P}_{\infty} \overline{\Omega}_{r}\right), \qquad r=1,2,3.
\end{equation}
The corresponding divisor $\mathcal{D}=P_1+P_2+P_3$  is non-special, and in the vicinity of $\mathcal{D}$
the map $\boldsymbol{\mathfrak{A}}$ is uniquely invertible.
\end{thm}

The rank of the Abel map is maximal on non-special divisors and decreases on special subvarieties of the Jacobian, the
{\it Wirtinger strata}. Here
$W^{(0)}= \overline{\boldsymbol{K}}_{\infty}\; \subset\;  W^{(1)}\;\subset\; W^{(2)}\; \subset\; \mathrm{Jac}(\mathcal{C})$ where
\begin{align}
\begin{split}
W^{(1)} :& \quad \boldsymbol{v} \in \mathrm{Jac}(X),\quad \boldsymbol{v}
=\int_{\infty}^P\overline{\boldsymbol{\Omega}}+\overline{\boldsymbol{K}}_{\infty}, \quad \forall P\in \mathcal{C},  \\
W^{(2)} :& \quad \boldsymbol{v}\in \mathrm{Jac}(X),\quad \boldsymbol{v}=\int_{\infty}^{P_1}\overline{\boldsymbol{\Omega}}+\int_{\infty}^{P_2}\overline{\boldsymbol{\Omega}}
+\overline{\boldsymbol{K}}_{\infty}, \quad \forall (P_1,P_2)\in \mathcal{C}\times\mathcal{C} \, .\end{split} \label{W_strata}
\end{align}

The equation $\theta(\boldsymbol{v};\tau)=0$ defines a codimension one
subvariety $\Theta\in \mbox{Jac} (\mathcal{C})$ (with singularities for $g>2$) called the {\it theta-divisor}, which
coincides with the stratum  $W^{(2)}$. This is equivalent to the fact that for
any points  $P_1$, $P_2\in \mathcal{C}$,
\begin{equation}
\theta\left( \int_{\infty}^{P_1} \overline{\boldsymbol{\Omega}} + \int_{\infty}^{P_2}\overline{ \boldsymbol{\Omega} }
+\overline{\boldsymbol{K}}_{\infty} ;\tau  \right) \equiv 0.  \label{KK}
\end{equation}
Note that a consequence of Riemann's theorem is that the vector $\overline{\boldsymbol{K}}_{\infty}$ for which
\eqref{KK} holds is unique.

We also introduce characteristics, represented by real $2\times3$ matrices
\[ [\varepsilon] = \left( \begin{array}{c}  \boldsymbol{\varepsilon}^T \\
{\boldsymbol{\varepsilon}'}^T   \end{array}\right)  =
\left( \begin{array}{ccc}  \varepsilon_1&\varepsilon_2&\varepsilon_3\\
\varepsilon_1'&\varepsilon_2'&\varepsilon_3'
 \end{array} \right) ,
\]
so that any vector $\boldsymbol{u}\in \mathrm{Jac}(\mathcal{C})$ can be written in the form
$\boldsymbol{u}= \mathcal{A}\boldsymbol{\varepsilon}'+
\mathcal{B}\boldsymbol{\varepsilon}, \quad \boldsymbol{\varepsilon}$,
$\boldsymbol{\varepsilon}' \in \mathbb{R}^3$.
We denote the characteristic of $\boldsymbol{u}$ by $[\boldsymbol{u}]$.
In what follows we concentrate on rational characteristics, in particular half-integer ones, for which
$\varepsilon_i,\varepsilon_j' = \frac12$ or $0$.
We shall also need the Riemann theta-functions with characteristics $[\varepsilon]$ given by
\begin{align*}
\theta \left[\boldsymbol{\varepsilon}^T \atop \boldsymbol{\varepsilon'}^T \right] (\boldsymbol{v};\tau)
&=\mathrm{exp} \left\{ \imath\pi
(\boldsymbol{\varepsilon}^T\tau\boldsymbol{\varepsilon}
+2\boldsymbol{\varepsilon}^T(\boldsymbol{v}+\boldsymbol{\varepsilon'}))\right\}
\theta(\boldsymbol{v}+\tau\boldsymbol{\varepsilon}+\boldsymbol{\varepsilon'};\tau)\\
& =\sum_{\boldsymbol{n}\in\mathbb{Z}^3}\mathrm{exp}
\left\{\imath\pi(\boldsymbol{n}+\boldsymbol{\varepsilon})^T\tau
            (\boldsymbol{n}+\boldsymbol{\varepsilon})
+2\imath\pi (\boldsymbol{n}+\boldsymbol{\varepsilon})^T(\boldsymbol{v}+\boldsymbol{\varepsilon'}) \right\},
\end{align*}

\subsection{Calculation of $\overline{\boldsymbol{K}}_{\infty}$.}
The vector of Riemann constant $\overline{\boldsymbol{K}}_{\infty}$ given by
\eqref{Riemann_K} includes Abelian integrals over
$\mathfrak{a}$-cycles, which are difficult to calculate directly.
It is known that if a curve has a point $P_\ast$ such that the canonical
divisor is linearly equivalent to $2(g-1)P_\ast$ then the vector of Riemann constants
with base point $P_\ast$ is a half-period \cite{fa73,fk80}. Such is the case for a hyperelliptic
curve.
For all $(n,s)$-curves Nakayashiki \cite{nak10} observed that these
admit a holomorphic differential that vanishes to order $2g-2$ at the point $P_0=(\infty,\infty)$ and so
we have the following.

\begin{prop} \label{K_trig}
The vector $\overline{\boldsymbol{K}}_{\infty}$ of the trigonal curve $\mathcal{C}$ is a half-period in
$\Jac (\mathcal{C})$.
\end{prop}

We may further restrict the choice of $\overline{\boldsymbol{K}}_{\infty}$. There are 64 half-periods in $\Jac (\mathcal{C})$, 28 odd and 36 even ones. Up to exponential terms the $\sigma$-function on $\mathcal{C}$ is
the $\theta$-function shifted by $\overline{\boldsymbol{K}}_{\infty}$ (see below).
Thus the leading terms of the expansion
of the $\theta$-function at $\overline{\boldsymbol{K}}_{\infty}$ coincides with those of the corresponding
$\sigma$-function. For $(3,4)$ curves it was shown in \cite{bel99} that the $\sigma$-function begins with an odd order
Schur function and it follows then that $\overline{\boldsymbol{K}}_{\infty}$ is an odd half-period.
Of course the explicit expression for $\overline{\boldsymbol{K}}_{\infty}$ depends
on the choice of homology basis on $\mathcal{C}$ and so to proceed further we must
first fix the homology basis. We may specify which half-period then corresponds to $\overline{\boldsymbol{K}}_{\infty}$ by first calculating the period matrix
and then checking the condition \eqref{KK} by computing the expansion of $\theta$-function near {\it all} odd
half-integer characteristics.
For example, for the curve $\mathcal{C}$ given by
\[  y^3-4yx^2-x^4-3x^3-4x^2-x+5=0\]
and in the Tretkoff--Tretkoff basis of cycles on it given by {\tt algcurves} package of {\tt Maple}
one obtains
\begin{equation} \label{char_K}
[\overline{\boldsymbol{K}}_{\infty}] =\frac12\left( \begin{array}{ccc}  1&1&0\\ 0&1&1 \end{array}  \right)\, .
\end{equation}

We note in passing that we already encounter here one of the differences with
Jacobians of hyperelliptic curves: for
a genus three {\it hyperelliptic} curve the vector of Riemann constants is given by an {\it even singular}
half-integer characteristic whereas the characteristic \eqref{char_K} is non-singular and odd.

\subsection{The $\sigma$-functions.} Apart from the theta-functions, in many cases it is more convenient to use
the $\sigma$-function of $\boldsymbol{u}$. To describe it, we first introduce the basis of 3 {\it meromorphic} differentials
$\boldsymbol{\Upsilon}=( \Upsilon_1, \Upsilon_2,\Upsilon_3 )^T$
having a unique pole at the infinity of $\mathcal{C}$ and specified by the pairing conditions
\begin{equation}
\Res\limits_{P=\infty} \;  \Omega_i(P)    \int_{P_0}^P  \Upsilon_j = \delta_{ij}, \qquad i,j=1,2,3.
\end{equation}
Then, following \cite{eemop07},
\begin{equation}
\boldsymbol{\Upsilon} = \frac{\mathrm{d}x}{f_y(x,y)} \left(
\begin{array}{c}
\Upsilon_1\\-2xy+\mu_1 x^2\\-x^2
\end{array}\right),
\end{equation}
where
\begin{align*}
\Upsilon_1&=-(5x^2+(\mu_1\mu_2-3\mu_3)x + \mu_2\mu_4+\mu_6)y+mu_2y^2 +3\mu_1x^3\\
&-(\mu_2^2+2\mu_3\mu_1-2\mu_4)x^2-(\mu_5\mu_2+\mu_6\mu_1+\mu_3\mu_4)x+\frac34\mu_1f_x(x,y)
-\left(\frac13\mu_2-\frac14 \mu_1^2\right)f_y(x,y) .
\end{align*}
The matrices of periods of the differentials
\begin{eqnarray}
\mathcal{S}=-\left(\oint_{a_k} \mathrm{d}\Upsilon_i\right)_{i,k=1,2,3},\quad
\mathcal{T}=-\left(\oint_{b_k} \mathrm{d}\Upsilon_i\right)_{i,k=1,2,3} ,
\end{eqnarray}
satisfy the generalized Legendre relation
\begin{equation}
\left(  \begin{array}{cc}   \mathcal{A}& \mathcal{B} \\
                             \mathcal{S}& \mathcal{T}   \end{array} \right) J    \left(  \begin{array}{cc}   \mathcal{A}& \mathcal{B} \\
                             \mathcal{S}& \mathcal{T}   \end{array} \right)^T=-2\pi\imath J, \qquad J=\left(  \begin{array}{cc}0_3&-1_3 \\
                             1_3&0_3     \end{array} \right) .
\end{equation}
Let us also introduce the normalized second period matrix $\varkappa = \mathcal{S}\mathcal{A}^{-1}$,
which is necessarily symmetric.

The fundamental $\sigma$-function of the curve $\mathcal{C}$ is defined by the formula (see e.g., \cite{bel97,eel00})
\begin{equation}
\sigma( \boldsymbol{u} ) = C\, \theta [\boldsymbol{K}] (\mathcal{A}^{-1} \boldsymbol{u} ) \, \mathrm{exp}
\left\{ \frac12 \boldsymbol{u}^T  \varkappa \,  \boldsymbol{u} \right\},
 \label{sigma}
\end{equation}
where $\theta [\boldsymbol{K}](\boldsymbol{v})$ is the theta-function with the characteristic corresponding to the vector of Riemann constants
(for a chosen base point and homology basis), and $C$ is a constant depending on the period matrix $\mathcal{A}$ and the
coefficients of the curve. The constant $C$ provides the modular invariance of (\ref{sigma}),
just as in the case of the Weierstrass elliptic $\sigma$-function.
An explicit expression for $C$ is given in \cite{eemop07} and it is not necessary for our exposition
(we will deal only with ratios of the sigma-function derivatives). According to the definition, the fundamental
 $\sigma$-function is normalized in such the way that its expansion at $\boldsymbol{u}\sim \boldsymbol{0}$
starts with the Schur-Weierstrass polynomial (see \cite{bel99} for details).
It follows that $\sigma(\boldsymbol{u})$ is just the theta-function of $\mathcal{C}$
whose rescaled argument is shifted by the vector $\overline{\boldsymbol{K}}_{\infty}$
and multiplied by a quadratic exponent of $\boldsymbol{u}$. Thus, the knowledge of the corresponding characteristic
$[\boldsymbol{K}]$ calculated above is important in the explicit description of $\sigma(\boldsymbol{u})$.

The function $\sigma( \boldsymbol{u} )$ is quasi-periodic just as is $\theta(\boldsymbol{v})$:
 when $\boldsymbol{u}$ is shifted  by a period
$\mathcal{A}\boldsymbol{n}+ \mathcal{B}\boldsymbol{m}$, $\boldsymbol{n},\boldsymbol{m}\in\mathbb{Z}^3$ then
  $\sigma( \boldsymbol{u} )$ is multiplied by the exponent
\begin{equation}
\sigma( \boldsymbol{u}+\mathcal{A}\boldsymbol{n}+ \mathcal{B}\boldsymbol{m} )
=\sigma( \boldsymbol{u} )\mathrm{exp}\left\{ \left(\mathcal{S}\boldsymbol{n}+  \mathcal{T}\boldsymbol{m}\right)^T\left( \boldsymbol{u}+\frac12\mathcal{A}\boldsymbol{n}+\frac12 \mathcal{B}\boldsymbol{m} \right)  \right\}
\end{equation}
For the case of the curve \eqref{trig}, at the origin $\boldsymbol{u}=0$,
the $\sigma$-function admits the following expansion (see \cite{eemop07})
\begin{gather} \label{sigma_exp}
\sigma( \boldsymbol{u} ) = u_1 - u_2^2 u_3+ \frac {1}{20} u_3^5 -\frac{\mu_2}{168}u_3^7+ \frac{\mu_2}{6} u_3^3 u_2^2
+\frac {\mu_3}{40}u_3^6 u_2-\frac{ \mu_3}{2}u_3^2 uï¿½_2^3 + \text{higher order terms} .
\end{gather}

It follows from  property \eqref{KK} and the definition \eqref{sigma} that for any points $P_1,P_2\in \mathcal{C}$
\begin{equation}
\sigma\left( \int_{\infty}^{P_1} \boldsymbol{\Omega} + \int_{\infty}^{P_2}\boldsymbol{\Omega} \right) \equiv 0.
\label{sig_0}
\end{equation}
That is, in $\boldsymbol{ u}$-coordinates on $\Jac(\mathcal{C})$, the stratum $W^{(2)}$ is given by the condition
$\sigma(\boldsymbol{u})=0$.

\subsection{Inversion of the Abel map in terms of the sigma function.}
We next introduce the Kleinian multi-index symbols
\begin{align}\begin{split}
\wp_{i,j}(\boldsymbol{u})&=-\frac{\partial^2}{\partial u_i\partial u_j} \;\mathrm{ln}\,\sigma(\boldsymbol{u}),
\qquad i,j=1,2,3,  \\
\wp_{i,j,k}(\boldsymbol{u})&
=-\frac{\partial^2}{\partial u_i\partial u_j\partial u_k}\;\mathrm{ln}\,\sigma(\boldsymbol{u}),
\qquad i,j,k=1,2,3  \\
&\vdots \end{split}\label{wp}
\end{align}
These are multiply periodic (Abelian) functions
\begin{equation}
\wp_{\mathcal{J}}( \boldsymbol{u} + \mathcal{A} \boldsymbol{n} + \mathcal{B} \boldsymbol{m} )
=\wp_{\mathcal{J}}(\boldsymbol{u}),
\end{equation}
where $\mathcal{J}$ is arbitrary multi-index with more than one entry.
It will also be convenient to denote throughout
$$\sigma_i (\boldsymbol{u})=\frac{\partial}{\partial u_i} \sigma(\boldsymbol{u} )$$
though these are not Abelian functions.

Then, as was first shown in \cite{bel00, eel00} (though in a somewhat hidden fashion),
the problem of inversion of the map \eqref{JIP} is reduced to solving the following two equations with respect to $x$ and $y$,
\begin{align}
&\wp_{33}(\boldsymbol{u})y+\wp_{23}(\boldsymbol{u})x+\wp_{13}(\boldsymbol{u})=x^2, \label{TJIP1}\\
&(\wp_{23}(\boldsymbol{u})-\wp_{333}(\boldsymbol{u}))y+(\wp_{22}(\boldsymbol{u})
-\wp_{233}(\boldsymbol{u}))x+\wp_{12}(\boldsymbol{u})-\wp_{133}(\boldsymbol{u})=2xy, \label{TJIP2}
\end{align}
the solutions of which give the coordinates of the points $P_1, P_2, P_3$ in \eqref{JIP}.
In contrast to the inversion problem for hyperelliptic case, equations (\ref{TJIP1}) and (\ref{TJIP2})
both contain the variables
$x$ and $y$. By elimination of $y$, one obtains a cubic equation with respect to $x$,
whereas elimination of $x$ yields a cubic equation for $y$. In the first case we get (the new expression,
though a simple consequence of the previous)
\begin{align}\begin{split}
\mathcal{P}(x;\boldsymbol{u})&=2x^3-( 3\wp_{23}(\boldsymbol{u})-\wp_{333}(\boldsymbol{u}) )x^2  \\
&-( \wp_{33}(\boldsymbol{u})\wp_{22}(\boldsymbol{u})+\wp_{23}(\boldsymbol{u})\wp_{333}(\boldsymbol{u})
-\wp_{33}(\boldsymbol{u})\wp_{233}(\boldsymbol{u})-\wp_{23}(\boldsymbol{u})^2+2\wp_{13} (\boldsymbol{u})) x \\
&-\wp_{33}(\boldsymbol{u})(\wp_{12}(\boldsymbol{u})-\wp_{133}(\boldsymbol{u}))
+\wp_{13}(\boldsymbol{u})(\wp_{23}(\boldsymbol{u})-\wp_{333}(\boldsymbol{u})) \\
&=2(x-x_1)(x-x_2)(x-x_3).\end{split}\label{Bolza3}
\end{align}
These formulae enable one to express the elementary symmetric functions of
$x_i$ as Abelian functions of $\boldsymbol u$. We shall use some of them in the next section.

\subsection{The inverse trace formula.}
Let $q_i =(0,y_i)$, $i=1,2,3$ be the three points on the curve $\mathcal{C}$ over $x=0$
and $\boldsymbol{r}_i= \int_\infty^{q_i}\boldsymbol{\Omega}$ be their images in $\Jac (\mathcal{C})$.
By applying the classical method of residues one can also obtain the following ``inverse trace formula'':
\begin{gather} \label{inv_trace}
\frac {1}{x_1} + \frac {1}{x_2} + \frac {1}{x_3}
= \sum_{i=1}^3 \partial_{ \mathcal{ V}_i } \log \sigma( \boldsymbol{r}_i- \boldsymbol{u} ) + k, \quad
\mathcal{V}_i= \left(\frac {1}{f_y(0,y_i)},0, \frac {y_i}{f_y(0,y_i)}\right)^T.
\end{gather}
Here $\mathcal{V}_i$ is the tangent
vector to $W^{(1)}\subset \Jac(\mathcal{C})$ at the point $\boldsymbol{r}_i$, and $k$ is a constant depending on the
curve $\mathcal{C}$ only.

Indeed, consider the single-valued function
$F(P)=\theta \left(\int_\infty^{P} \overline{ \boldsymbol{\Omega}} - {\boldsymbol v}-
\overline{\boldsymbol{K}}_{\infty} \right)$
defined on a simply connected dissection $\widetilde{ \mathcal{C}}$ of $\mathcal{C}$ with  boundary $\partial \widetilde{ \mathcal{C}}$.
Then, for a meromorphic function $f(P)$, $P\in \mathcal{C}$ with the poles $Q_1,\dots, Q_s$ on $\mathcal{C}$,
the residue formula gives (see, e.g., \cite{dub81},\cite{bbeim94})
\begin{equation} \label{residue}
f(P_1)+ f(P_2)+f(P_3) = \frac{1}{2\pi \imath} \oint\limits_{\partial \widetilde{ \mathcal{C}}} f(P) d \log F(P) -
\sum_{k=1}^s \Res\limits_{Q_s} f(P) d \log F(P).
\end{equation}
Set here $f(P)=1/x$ and observe that this function has simple poles precisely
at $Q_i=q_i$, $i=1,2,3$ and that the expression is independent of the choice of local parameter.  Then, using the expansions of $F(P)$ in the neighborhood of $q_i$ and the
relation \eqref{sigma} between $\theta(\boldsymbol{v})$ and $\sigma (\boldsymbol{u})$, one arrives at \eqref{inv_trace}.
Alternatively, one can derive a $\sigma$-function analogue of formula \eqref{residue}.

Notice that $\mathcal{V}_1 + \mathcal{V}_2 + \mathcal{V}_3=0$, which, in view of the quasi-periodic property of
$\sigma (\boldsymbol{u})$, ensures that the right hand side of \eqref{inv_trace}
remains unchanged when $\boldsymbol{u}$ changes by a period vector of $\Jac(\mathcal{C})$.
The constant $k$ can be calculated explicitly by letting  $x_1,x_2,x_3 \to\infty$ in \eqref{inv_trace} and evaluating the right hand side (see also formula \eqref{inv_tr2} below).

For the curve relevant to the Goryachev case under consideration we have simplifications.

\begin{prop} \label{inv_sum} In the special case $\mu_8=0, \mu_{12}=1$ in the equation of the trigonal curve \eqref{trig} one has
\begin{equation} \label{sp_case}
q_i= (0, \rho^{i}), \quad \rho= \exp (2\pi \imath /3), \quad
\mathcal{V}_i= \frac 13 \left( \frac {1}{\rho^{2i}}, 0, \frac {1}{\rho^i } \right)
= \frac 13 ( \rho^i, 0, \rho^{2i} ).
\end{equation}
In this case formula \eqref{inv_trace} takes the form
\begin{gather} \label{inv_tr2}
\frac {1}{x_1} + \frac {1}{x_2} + \frac {1}{x_3}
= \frac 13  \sum_{i=1}^{3} \left(
\rho^{i} \partial_1 \log \sigma(\boldsymbol{r}_{i}-\boldsymbol{u}) +
\rho^{2 i} \partial_3 \log \sigma(\boldsymbol{r}_{i}- \boldsymbol{u} )\right)+ k, \\
k = -\frac 13 \left( \frac{\sigma_{23}(\boldsymbol{r}_3)}{\sigma_2(\boldsymbol{r}_3) }+
\rho^2 \frac{\sigma_{23}(\boldsymbol{r}_1)}{\sigma_2(\boldsymbol{r}_1)}
+ \rho^4 \frac{\sigma_{23}(\boldsymbol{r}_2)}{\sigma_2(\boldsymbol{r}_2) }\right)\, , \quad
 \boldsymbol{r}_{i}=\int_{\infty}^{\rho_{i}} \boldsymbol{\Omega}.
\label{k0}
\end{gather}
\end{prop}

{\it Proof} of Proposition \ref{inv_sum}.  The formula \eqref{inv_tr2} follows from \eqref{inv_trace}
under the conditions \eqref{sp_case}.
To calculate the constant $k$, let the points $P_1, P_2, P_3\in \mathcal{C}$ to tend to $\infty$ in such a way that
under the Abel map \eqref{JIP'} one has $u_1 \equiv u_3\equiv 0$, and the coordinate $u_2$ tends to zero.
Then the left hand of \eqref{inv_tr2} tends to zero,
while the right hand side becomes the sum of $k$ and of the terms
$$
\dfrac{ \rho^{2 i } \sigma_1(\boldsymbol{r}_i ) + \rho^{i}
\sigma_3 ( \boldsymbol{r}_i )}{3 \sigma ( \boldsymbol{r}_i )}\, .
$$
Now as $\sigma_1(\boldsymbol{r}_i)=\sigma_3(\boldsymbol{r}_i)=\sigma(\boldsymbol{r}_i)=0$
this expression in indeterminate, but
applying l'Hopital's rule to each term we obtain
\begin{align}\begin{split}
\lim\limits_{u_2 \to 0}
\frac{ \rho^{2i} \sigma_1( \boldsymbol{r}_i + u_2 \boldsymbol{e}_2 )+ \rho^{i}
\sigma_3 ( \boldsymbol{r}_i + u_2\boldsymbol{e}_2 )}{3 \sigma ( \boldsymbol{r}_3+ u_2 \boldsymbol{e}_2 ) }
& = \lim\limits_{u_2 \to 0} \frac{\rho^{2i}  \sigma_{12}(\boldsymbol{r}_i + u_2 \boldsymbol{e}_2 )
+ \rho^{i} \sigma_{23}(\boldsymbol{r}_i +u_2 \boldsymbol{e}_2 )}
{3\sigma_2(\boldsymbol{r}_i + u_2 \boldsymbol{e}_2 ) }
 \\
& =\frac{\rho^{2i}  \sigma_{12}(\boldsymbol{r}_i )
+ \rho^{i} \sigma_{23}(\boldsymbol{r}_i )}{3 \sigma_2(\boldsymbol{r}_i) },
 \end{split} \label{Hopital1}
\end{align}
where $\boldsymbol{e}_2=(0,1,0)^T$. The latter fractions are well determined.
Finally, as we will prove in Proposition 4.2 below, $\sigma_{12}(\boldsymbol{r}_i )=0$ for any $i=1,2,3$,
and so \eqref{Hopital1} gives \eqref{k0}. $\square$



\section{Solving the inversion problem on the stratum $W^{(2)}$ and sigma-function solutions of the Goryachev system}
We now identify the trigonal curve  \eqref{curve1} that appears in the quadratures for the Goryachev system
and the curve $\widetilde{\mathcal{C}}$ in \eqref{trig} by setting
\begin{equation}
\mu_1=\mu_4=0, \quad \mu_2=-2h_1/a, \quad \mu_5=\mu_8=0, \quad
\mu_3=4h_2\left(\frac{2}{a}\right)^{2/3}, \quad
\mu_6=-2b/a,\quad \mu_9=0, \quad \mu_{12}=1. \label{mu's}
\end{equation}
Notice again that the map $\mathcal{C}\times \mathcal{C} \to \Jac(\mathcal{C})$ in \eqref{AM2} contains only 2 points on $\mathcal{C}$
and cannot be identified with the full
Abel map \eqref{JIP}. According to the previous observations, \eqref{AM2} maps the symmetric product $\mathcal{C}\times \mathcal{C}$
to the stratum $W^{(2)}\subset \Jac (\mathcal{C})$ given analytically by the condition $\sigma(\boldsymbol{u})=0$.

\subsection{The inversion of the Abel map on the stratum $W^{(2)}$.}
One can extend this map by adding a third {\it fixed} point on $\mathcal{C}$, in particular, one of the three points
$q_i=(0, \rho^{i})\in \mathcal{C}$:
$$
 \int_{2\infty}^{P_1+P_2} \Omega + \int_\infty^{q_i} \Omega = \boldsymbol{u} + {\boldsymbol r}_i, \qquad
{\boldsymbol r}_i =\int_\infty^{q_i} \Omega ,
$$
$\boldsymbol{u}$ being the right hand side of \eqref{AM2}.
Then, by using \eqref{Bolza3}, one obtains
the following formal complex expressions for the symmetric functions of $x_1, x_2$
(which hold for any $i=1,2,3$ on the right hand side):
\begin{align}
x_1+x_2 & = 3\wp_{23}( \boldsymbol{u} +\boldsymbol{r}_{i} ) -\wp_{333}(\boldsymbol{u}+ \boldsymbol{r}_{i}), \label{sumr} \\
x_1 x_2 & = \wp_{33}(\boldsymbol{u}+ \boldsymbol{r}_{i} )\wp_{22}(\boldsymbol{u}+\boldsymbol{r}_{i} )
+\wp_{23}(\boldsymbol{u}+ \boldsymbol{r}_{i} )\wp_{333}(\boldsymbol{u}+ \boldsymbol{r}_{i} ) \notag \\
& \quad -\wp_{33}( \boldsymbol{u} + \boldsymbol{r}_{i} )\wp_{233}(\boldsymbol{u}+ \boldsymbol{r}_{i} )
-\wp_{23}^2(\boldsymbol{u}+\boldsymbol{r}_{i} ) +2\wp_{13} (\boldsymbol{u}+ \boldsymbol{r}_{i} ) , \label{prod_2}
\end{align}
where, according to \eqref{AM2}, $u_3=- 2\imath t_1/3, \; u_2=-2\imath t_2/3$, and the coordinate
$u_1$ is defined (but not uniquely!) from the transcendental condition $\sigma (\boldsymbol{u})=0$.

On the other hand, letting $x_3\to \infty$ in the inverse trace formula \eqref{inv_tr2}, we find
\begin{gather*}
 \frac{x_1+x_2}{x_1 x_2}
= \frac 13  \sum_{i=1}^{3} \left(
\rho^{i} \partial_1 \log \sigma(\boldsymbol{r}_{i}-\boldsymbol{u}) +
\rho^{2i} \partial_3 \log \sigma(\boldsymbol{r}_{i}- \boldsymbol{u} )\right)
+ k \bigg |_{\sigma (\boldsymbol{u}) =0}, \\
k = -\frac 13 \sum_{i=1}^{3}\frac{\rho^{2i}\sigma_{2,3}(\boldsymbol{r}_{i})}
{\sigma_{2}(\boldsymbol{r}_{i})} .
\end{gather*}
Upon combining the above with \eqref{sumr}, one obtains the following alternative to \eqref{prod_2}
\begin{equation} \label{x12}
x_1 x_2  = \frac {3\wp_{23}( \boldsymbol{u} + \boldsymbol{r}_{i} ) -\wp_{333}(\boldsymbol{u}+ \boldsymbol{r}_{i} )}
{\sum_{i=1}^3 \partial_{ \mathcal{V}_i } \log \sigma( \boldsymbol{u} - \boldsymbol{r}_{i})
+ k } \bigg |_{\sigma (\boldsymbol{u}) =0} \, , \quad
\mathcal{V}_i= \frac 13 ( \rho^i, 0, \rho^{2i} ).
\end{equation}

There is another way of writing the formal solution to the inversion of \eqref{AM2}.
(For the case of hyperelliptic curves this was proposed in \cite{gr90,jo92} and also used in
\cite{EPR03}.) Namely, consider again the full Abel map \eqref{JIP} with the three points
$P_1=(x_1,y_1), P_2=(x_2,y_2), P_3 =(x_3, y_3)$ and observe that
$$
x_1+x_2 = \lim_{x_3\to\infty} \frac{x_1x_2+x_2 x_3+x_3 x_1}{x_1+x_2+x_3} .
$$
Then, in view of \eqref{Bolza3},
$$
x_1+x_2 = \lim_{P_3\to\infty} \frac{\wp_{33}( \boldsymbol{u} )\wp_{22}(\boldsymbol{u})+\wp_{23}(\boldsymbol{u})\wp_{333}(\boldsymbol{u})
-\wp_{33}(\boldsymbol{u})\wp_{233}(\boldsymbol{u})-\wp_{23}(\boldsymbol{u})^2+2\wp_{13}(\boldsymbol{u})}
{ 3\wp_{23}(\boldsymbol{u})-\wp_{333} (\boldsymbol{u}) }.
$$
Using the definition of $\wp_{ij}$ in \eqref{wp} and taking the limit (for which $\sigma(\boldsymbol{u})=0$)
one obtains
\begin{equation} \label{symm1}
x_1+x_2 =\frac12 \phi(\boldsymbol{u})\frac{\sigma_3(\boldsymbol{u})\sigma_{2,3}
(\boldsymbol{u}) -\sigma_2(\boldsymbol{u})\sigma_{3,3}(\boldsymbol{u})}{\sigma_3^3(\boldsymbol{u})}
+\frac12\frac{\sigma_2(\boldsymbol{u})\phi_3(\boldsymbol{u})-\sigma_3(\boldsymbol{u})\phi_2(\boldsymbol{u})}
{\sigma_3^3(\boldsymbol{u})},
\end{equation}
where we set
\begin{equation} \label{phis'}
\phi(\boldsymbol{u})= \sigma_2(\boldsymbol{u})-\sigma_{3,3}(\boldsymbol{u}),\quad
\phi_j(\boldsymbol{u})=\sigma_{j,2}(\boldsymbol{u})-\sigma_{j,3,3}(\boldsymbol{u}), \quad j=1,2,3 , \quad \boldsymbol{u} \in W^{(2)},
\end{equation}
and $\boldsymbol u$ is given by the right hand side of \eqref{AM2}.
In a similar way we find
\begin{equation}\label{symm2}
x_1x_2=\frac12 \phi(\boldsymbol{u})\frac{\sigma_1(\boldsymbol{u})\sigma_{3,3}(\boldsymbol{u})
-\sigma_3(\boldsymbol{u})\sigma_{1,3}(\boldsymbol{u})}{\sigma_3^3(\boldsymbol{u})}
+\frac12\frac{\sigma_3(\boldsymbol{u})\phi_1(\boldsymbol{u})-\sigma_1(\boldsymbol{u})
\phi_3(\boldsymbol{u})}{\sigma_3^3 (\boldsymbol{u})}, \qquad \boldsymbol{u} \in W^{(2)}.
\end{equation}

Next, using the relation (\ref{TJIP1}) for the pairs $(x_1,y_1), (x_2, y_2)$, we find
\begin{equation} \label{TJY}
y_j=(x_j^2-\wp_{23}(\boldsymbol u) x_i-\wp_{13}(\boldsymbol u))/\wp_{33}(\boldsymbol u),
\end{equation}
and, therefore,
$$
\frac{x_2y_1-x_1y_2}{x_2-x_1}
=\frac{x_1x_2-\wp_{13}(\boldsymbol{u})}{\wp_{33}(\boldsymbol{u})}, \quad
\frac{y_1-y_2}{x_2-x_1} =\frac{x_1+x_2-\wp_{23}(\boldsymbol{u})}{\wp_{33}( \boldsymbol{u} )} ,
$$
where $ \boldsymbol u$ is again given by the right hand side of \eqref{AM2}.
In view of \eqref{wp} and the condition $\sigma (\boldsymbol{u})=0$, this gives the following new compact expressions
\begin{equation}\label{sigma_quo}
\frac{x_2y_1-x_1y_2}{x_2-x_1}
=\left. \frac{\sigma_1(\boldsymbol{u})}{\sigma_3(\boldsymbol{u})}\right|_{ \sigma(\boldsymbol{u})=0 }, \qquad
\frac{y_1-y_2}{x_2-x_1}
=\left. \frac{\sigma_2(\boldsymbol{u})}{\sigma_3( \boldsymbol{u} )} \right|_{\sigma(\boldsymbol{u})=0} .
\end{equation}
The above formulae lead to the following analytic description of the Wirtinger strata defined in \eqref{W_strata}.

\begin{prop} \label{sigmas_on_W}
\begin{enumerate}
\item The strata $W^{(0)},W^{(1)},W^{(2)} $ are given by the conditions
\begin{align}
&W^{(0)}: \quad \sigma(\boldsymbol{u})=\sigma_3(\boldsymbol{u})=\sigma_2(\boldsymbol{u})=0, \notag \\
&W^{(1)}:\quad  \left\{ \boldsymbol{u}\vert\;\; \sigma( \boldsymbol{u} )=0,\; \sigma_3(\boldsymbol{u})=0   \right\},
\label{JIPTRIG}\\
&W^{(2)}:\quad  \left\{ \boldsymbol{u}\vert\;\; \sigma(\boldsymbol{u})=0   \right\} \notag
\end{align}

\item  The coordinates $x,y$ of the curve $\mathcal{C}= W^{(1)} \subset \Jac (\mathcal{C})$ admit the parameterization
\begin{equation} \label{x_y_1}
x_1 = - \frac{\sigma_1(\boldsymbol{u})}{\sigma_2(\boldsymbol{u})}, \quad
y_1 = \frac{ \sigma_1 (\boldsymbol{u}) \sigma_{2,3} (\boldsymbol{u})
-\sigma_2 (\boldsymbol{u}) \sigma_{1,3} (\boldsymbol{u}) } {\sigma_2^2(\boldsymbol{u})}, \quad
\boldsymbol{u} \in W^{(1)}.
\end{equation}
\end{enumerate}
\end{prop}

\noindent{\it Proof.} The description of $W^{(2)}$ was already given by \eqref{sig_0}.
To pass to the stratum $W^{(1)}$ one should let $x_2\rightarrow \infty$.
Then both sides of (\ref{symm1}) and (\ref{symm2}) tend to infinity, which happens if and only if
$\sigma(\boldsymbol{u})=0$ and $\sigma_3(\boldsymbol{u})=0$.

Next, in view of (\ref{symm1}), (\ref{symm2}), and the condition $\sigma_3(\boldsymbol{u})=0$,
$$ 
\lim_{x_2\to\infty} \frac{x_1x_2}{x_1+x_2}= x_1 =- \frac{\sigma_1(\boldsymbol{u})}{\sigma_2(\boldsymbol{u})}, \quad
\boldsymbol{u}=\int_{\infty}^{P_1} \boldsymbol{\Omega}\, ,
$$ 
which gives the first expression in \eqref{x_y_1}.
Letting here $x_1\to \infty$ gives $\sigma_2( \boldsymbol{u} )=0$ for $\boldsymbol{u}=0$. The latter also follows
directly from the expansion \eqref{sigma_exp}. The second formula \eqref{x_y_1} is obtained by the appropriate limit from
\eqref{TJY}. $\square$

We shall also use the following

\begin{prop} \label{sigmaexp}
Let ${\bf w}=\boldsymbol{u}- \boldsymbol{r}_i=(w_1,w_2,w_3)$ and set
$\sigma_{j,\dots,k}^{(i)} =\sigma_{j,\dots,k} ( \boldsymbol{r}_i)$.
Then \\ $\sigma_1^{(i)}=\sigma_3^{(i)}=0$ and
the first few coefficients of the expansion near $\boldsymbol{r}_i \in W^{(1)}$, $i=1,2,3$
\begin{gather}\begin{split}
\sigma (\boldsymbol{r}_i + {\bf w})= \sigma_2^{(i)} \, w_{2}
+ \frac {1}{2} \,\sigma_{1, \,1}^{(i)} \,{w_{1}}^{2} + \sigma_{1, \,3}^{(i)} \,w_1\,w_3 +
\frac {1}{2} \, \sigma_{3, \,3}^{(i)}  \,w_3^{2} +
 \frac {1}{2} \, \sigma_{2, \,2}^{(i)}  \,w_2^{2} +
\sigma_{1, \,2}^{(i)} \,w_{1}\,w_{2} + \sigma _{2, \,3}^{(i)} \, w_{2}\,w_{3}   \\
+ \frac {1}{6} \, \sigma_{1, \,1, \,1}^{(i)} \,{w_{1}}^{3}
+ \frac {1}{2} \,\sigma_{1, \,1, \,2}^{(i)} \,{w_{2}}\,{w_{1}}^{2}
+ \frac {1}{2} \,\sigma_{1, \,1, \,3}^{(i)} \,{w_{3}}\,{w_{1}}^{2} +
\frac {1}{2} \,\sigma_{1, \,2, \,2}^{(i)} \,{w_{1}}\,{w_{2}}^{2} +
\sigma_{1, \,2,\,3}^{(i)} \,{w_{3}}\,{w_{1}}\,{w_{2}} \\
+ \frac {1}{2} \,\sigma_{1, \,3, \,3}^{(i)}
\,w_{3}^2\,w_{1} + \frac {1}{6} \,\sigma_{2, \,2, \,2}^{(i)} \,w_2^{3}
+ \frac {1}{2} \,\sigma_{2, \,2, \,3}^{(i)} \,w_{3}\,w_{2}^{2} +
\frac {1}{2} \,\sigma_{2, \,3, \,3}^{(i)} \,w_{3}^2\,w_{2} +
\frac {1}{6} \,\sigma_{3, \,3, \,3}^{(i)} \,w_3^{3} + O(w^4)\end{split} \label{sigma_r}
\end{gather}
are related as follows
\begin{gather} \label{s_sr}
\sigma_{3,3}^{(i)} = \sigma_2 ^{(i)} , \quad  \sigma_{1,3}^{(i)}=-\rho^i \sigma_2^{(i)}, \quad
\sigma_{1,1}^{(i)} = - \frac{2}{\rho^i}\sigma_2^{(i)},  \quad
\sigma_{3,3,3}^{(i)} = 3\sigma_{2,3}^{(i)}  , \\
\sigma_{1,2}^{(i)} = \sigma_{1,1,1}^{(i)}=0, \quad
\sigma_{1,1,3}^{(i)} =\sigma_{1,3,3}^{(i)} = -2\rho^{i}  \sigma_{2,3}^{(i)}.  \label{ders_rho}
\end{gather}
Here, as above, $\rho= \exp(2\pi \imath/3)$.
\end{prop}

The proof uses the expansions of the coordinate $y$ near the points $q_i=(0,\rho^i)\in \mathcal{C}$,
\begin{align} \begin{split}
y & = \rho^i + \frac 13 (\mu_2(\rho^i +1) + \mu_6 \rho^i ) x^2 + \frac{\mu_3}{3} \rho^i x^3 + O(x^4)
\\
 & = \rho^i + \frac 13 (-\mu_2 \rho^{2i} + \mu_6 \rho^i) x^2 + \frac{\mu_3}{3} \rho^i x^3 + O(x^4). \end{split}
\label{y_r}
\end{align}
and the corresponding expansions of the holomorphic differentials \eqref{difs_t}. Thus, for example, near $(0,\rho)$ one has
\begin{align}
\Omega_{1} & =
\left(  \frac {\rho }{3}  - \frac {1}{9}\,{\displaystyle \frac {2\,\rho ^{2}\,{\mu _{6}} - {\mu _{2}}}{
\rho }} \,x^{2} - \frac {2}{9} \,\rho \,\mu _{3}\,x^{3} + {\displaystyle \frac {1}{27}} \,
{\displaystyle \frac {(3\,\rho ^{2}\,{\mu _{6}} - 2\,{\mu _{2}})\,{\mu _{6}}}{\rho }}
\,x^{4} 
+ O(x^5) \right) dx, \notag \\
\Omega_{2} &=
\left ( {\displaystyle \frac {\rho }{3}} \,x - {\displaystyle \frac {1}{9
}} \,{\displaystyle \frac {2\,\rho ^{2}\,{\mu _{6}} - {\mu _{2}}
}{\rho }} \,x^{3} - {\displaystyle \frac {2}{9}} \,\rho \,{\mu _{
3}}\,x^{4} + {\displaystyle \frac {1}{27}} \,{\displaystyle
\frac {(3\,\rho ^{2}\,{\mu _{6}} - 2\,{\mu _{2}})\,{\mu _{6}}}{\rho }} \,x^{5}+O(x^6)\right) dx, \label{exp_Om_rho} \\
\Omega_{3} &=
\left( {\displaystyle \frac {\rho ^{2}}{3}}  - {\displaystyle \frac {1}{
9}} \,\rho ^{2}\,{\mu _{6}}\,x^{2} - {\displaystyle \frac {1}{9}
} \,{\mu _{3}}\,\rho ^{2}\,x^{3} + {\displaystyle \frac {1}{27}}
\,{\displaystyle \frac {\rho ^{4}\,{\mu _{6}}^{2} + \rho ^{2}\,{
\mu _{6}}\,{\mu _{2}} - {\mu _{2}}^{2}}{\rho ^{2}}} \,x^{4} +
 O(x^5) \right) dx  \notag
\end{align}
and the expansions near $(0,\rho^2), (0,1)$ are obtained by replacing $\rho$ above by $\rho^2$ and 1 respectively.

The remainder of the proof is technical and is presented in the Appendix.

\subsection{Analytic properties of the inversion on $W^{(2)}$.}
One should stress that formulae \eqref{symm1}-\eqref{sigma_quo}
provide only a local analytic solution to the inversion of the incomplete Abel map \eqref{AM2}. This is because
$\sigma( \boldsymbol{u})=0$ is a transcendental equation and
the argument $u_1$ is an infinitely-valued complex function of $u_2, u_3$.
This fact also admits a geometric description. Namely,
let $\mathcal{W}^{(2)} \subset {\mathbb C}^3(u_1,u_2,u_3)$ be the universal covering of the stratum $W^{(2)}\subset \Jac(\mathcal{C})$ and consider the projection
$\pi\, : \, \mathcal{W}^{(2)} \to {\mathbb C}^2 (u_2, u_3)$.

\begin{prop} \label{proj_strata} Assume that $\mathrm{Jac}(\mathcal{C})$ has no Abelian subvarieties. Then,
under the projection $\pi$, the variety $\mathcal{W}^{(2)}$ is an infinitely-sheeted covering of
${\mathbb C}^2 (u_2, u_3)$ ramified along the subvariety $\bar W \subset \mathcal{W}^{(2)}$ defined by the
conditions $\{\sigma(u)=0, \; \sigma_1(u) =0\}$.

Moreover, let $\{ \boldsymbol{u}^* \}\subset \mathcal{W}^{(2)}$ be the equivalence class corresponding to
any point $\boldsymbol{u}^*\in W^{(2)}$. Then the projection $\pi \{ \boldsymbol{u}^* \}$ forms
a dense set on ${\mathbb C}^2 (u_2, u_3)$.
\end{prop}

Note that in the case of hyperelliptic curves of genus 2, when the codimension one stratum $W^{(1)}$
coincides with the curve itself, a similar description was made by Jacobi in connection with
the inversion of a single hyperelliptic integral (see \cite{marku92}), whereas for hyperelliptic curves of any genus
and strata of any codimension a similar theorem was proven in \cite{abendfed00}.
\medskip

\noindent{\it Proof of Proposition \ref{proj_strata}.} The proof follows the same lines as that in \cite{abendfed00}.
Namely, let $\mathfrak v_1,\dots, \mathfrak v_6 \in {\mathbb C}^2 (u_2, u_3)$ be the $\pi$-projections
of six independent period vectors of $\Jac(\mathcal{C})$. For any point $\boldsymbol{u}^* \in W^{(2)}$, the projections of its equivalence
class on $\mathcal{W}^{(2)}$ has the form
$$
 \left \{ \pi (\boldsymbol{u}^*)+ \sum_{j=1}^6 m_j {\mathfrak v}_j \mid m_j \in {\mathbb Z} \right\} .
$$
Under the condition of the proposition the periods themselves are not commensurable and consequently
the integer coefficients $m_j$ can always be chosen in such a way that the above sum
will give a point in any small neighborhood of any point of ${\mathbb C}^2 (u_2, u_3)$ fixed \emph{a priori}.
Next, since $\mathcal{W}^{(2)}$ is defined by the transcendental equation $\sigma(\boldsymbol{u})=0$, for any
point $(u_2^*,u_3^*)\in {\mathbb C}^2$ there exists an infinite number of solutions $u_1$. In other words,
the covering $\pi\, : \, \mathcal{W}^{(2)} \to {\mathbb C}^2 (u_2, u_3)$ has an infinite number of sheets.
Finally, by the implicit function theorem, $\pi$ is ramified over the points satisfying $\sigma_1(u)=0$.
This establishes the proposition. $\square$

\subsection{Sigma-function solutions of the Goryachev system.}

We conclude the section with the formal $\sigma$-function solutions for the original variables of the Goryachev system.
First, applying the transformation (\ref{subs}) to \eqref{exp_lambdas}, one gets
\begin{equation} \label{solutioins_x_y}
\begin{aligned}
\gamma_3^{2} &= \left( \frac{x_2y_1-x_1y_2}{x_2-x_1} \right)^3, \quad
J_3= \frac{3}{2} \sqrt{\frac{a}{2}} \frac{y_1-y_2}{x_2-x_1}, \\
\gamma_2+ \imath \gamma_1
& = \frac {\imath}{x_1 x_2} \left(1 -\left( \frac{x_2y_1-x_1y_2}{x_2-x_1} \right)^3 \right), \quad
\gamma_2 -\imath \gamma_1 = -\imath \, x_1 x_2, \\
J_1 + \imath J_2 & = -\sqrt{\frac a 2} (x_1+x_2) \left( \frac{x_2y_1-x_1y_2}{x_2-x_1} \right)^{-1/2}, \\
J_1 - \imath J_2 & = \frac{1}{\gamma_2-\imath \gamma_1}
\left((J_1+\imath J_2) (\gamma_2+\imath\gamma_1)+ 2 \imath J_3\gamma_3 \right) .
\end{aligned}
\end{equation}
Then, upon comparing the above with the sigma-function expressions \eqref{symm1}, \eqref{symm2}, \eqref{sigma_quo},
one obtains after simplification
\begin{align}
\gamma_3^{2/3} & =\frac{\sigma_1(\boldsymbol{u}) } {\sigma_3(\boldsymbol{u} ) }
\bigg |_{ \sigma(\boldsymbol{u})=0 } , \quad
J_3= \frac{3}{2} \sqrt{ \frac a 2} \frac{\sigma_2(\boldsymbol{u}) } {\sigma_3(\boldsymbol{u}) }
\bigg |_{\sigma(\boldsymbol{u})=0 }, \notag \\
  \gamma_2 -\imath \gamma_1 & = -\frac {\imath}{2} \frac{\Phi_2(\boldsymbol{u})}
 { \sigma_3^3(\boldsymbol{u}) } \bigg |_{ \sigma(\boldsymbol{u})=0 }  , \notag \\
 \gamma_2+ \imath \gamma_1 &
= 2 \imath \frac{ \sigma_3^3(\boldsymbol{u}) - \sigma_1^3(\boldsymbol{u}) } {\Phi_2(\boldsymbol{u})}\bigg |_{ \sigma(\boldsymbol{u})=0 }
 , \label{sols_x} \\
J_1 + \imath J_2 & = - \sqrt{\frac a 2}\frac 12 \frac{\Phi_1 (\boldsymbol{u})}
{\sqrt{ \sigma_1(\boldsymbol{u}) \sigma_3(\boldsymbol{u}) } \,\sigma_3^2(\boldsymbol{u}) }\bigg |_{ \sigma(\boldsymbol{u})=0 }
 , \notag \\
J_1 - \imath J_2 & =\sqrt{\frac a 2} \, \frac{\sqrt{ \sigma_3(\boldsymbol{u})} } {\sqrt{\sigma_1(\boldsymbol{u})} }
\, \frac{\Phi_1 (\boldsymbol{u}) ( \sigma_3^3 (\boldsymbol{u}) - \sigma_1^3(\boldsymbol{u}))
+3 \Phi_2(\boldsymbol{u})\sigma_1^2(\boldsymbol{u}) \sigma_2(\boldsymbol{u})) } {\Phi_2^2(\boldsymbol{u}) } \bigg |_{ \sigma(\boldsymbol{u})=0 },
\notag
\end{align}
where we set
\begin{align*}
\Phi_1 (\boldsymbol{u}) & = \phi(\boldsymbol{u}) \left( \sigma_3(\boldsymbol{u})\sigma_{2,3}
(\boldsymbol{u}) -\sigma_2(\boldsymbol{u})\sigma_{3,3}(\boldsymbol{u}) \right)
+\sigma_2(\boldsymbol{u})\phi_3(\boldsymbol{u})-\sigma_3(\boldsymbol{u})\phi_2(\boldsymbol{u}), \\
\Phi_2 (\boldsymbol{u}) & = \phi(\boldsymbol{u})\left( \sigma_1(\boldsymbol{u})\sigma_{3,3}(\boldsymbol{u})
-\sigma_3(\boldsymbol{u})\sigma_{1,3}(\boldsymbol{u}) \right)
+ \sigma_3(\boldsymbol{u})\phi_1(\boldsymbol{u})-\sigma_1(\boldsymbol{u})
\phi_3(\boldsymbol{u}) .
\end{align*}
Here $\phi(\boldsymbol{u})$ and its derivatives were defined in \eqref{phis'}, and the components
$u_2, u_3$ of $\bf u$ are linear functions of $t_1, t_2$ as described in \eqref{AM2}. These expressions give the
solution of the Goryachev system for a general flow under the Hamiltonians $H_1$, $H_2$.

\section{Expansions of the functions $x_i, y_i$ and the Painlev\'e analysis of the system.}

Apart from the formal $\sigma$-function solutions \eqref{sols_x}
it is important to know the complex singularities of the functions $J_i(t), \gamma_i(t)$: their
poles, order of branching, etc. Here we shall make connection with the Painlev\'e analysis of \cite{AvM1}.
For simplicity we concentrate on the complex flow generated by the quadratic Hamiltonian $H_1$ (time $t=t_1$).
As we have seen, this flow lies on the stratum $W^{(2)}\subset \Jac (\mathcal{C})$, on which the coordinate
$u_1$ is a transcendental function of $u_2,u_3$ and for which $u_2$ is constant along the flow.

It follows from \eqref{sols_x} that most of the variables have poles when the $u_3$-flow (i.e. the $t_1$-flow)
on $W^{(2)}$ crosses the substratum $W^{(1)}$, on which $\sigma_3(\boldsymbol{u})=0 $. Also, the right hand sides of
\eqref{sols_x} may have branching only when the coordinate $u_1$ (as a solution of $\sigma(u_1,u_2, u_3)=0$)
ceases to be a locally meromorphic function of $u_2,u_3$ (and, therefore, of $t_1$).
As observed above, the implicit function theorem means this occurs along the 1-dimensional analytic subvariety
$\mathcal{Z}=\{ \sigma(\boldsymbol{u})=0, \sigma_1(\boldsymbol{u})=0\} \subset W^{(2)}$.
(The solutions for $J_2 \pm \imath J_1$ in \eqref{sols_x} have additional branching along $W^{(1)}$
due to presence of square roots.)

To describe the local behavior of $u_1=u_1(u_2, u_3)$ in detail, choose a point $\boldsymbol{u}_0 \in W^{(1)}$.
Then $\sigma(\boldsymbol{u}_0 )= \sigma_3(\boldsymbol{u}_0)=0$.
Let $\delta u_i$ be the increments of the coordinates $u_i$ such that
$\boldsymbol{u}_0 + \delta\boldsymbol{u} \in W^{(2)}$. Then the following expansion holds
\begin{equation} \label{sigma_exp_rho}
\sigma(\boldsymbol{u}_0 + \delta\boldsymbol{u} )
= \sigma_1( \boldsymbol {u}_0 )\, \delta u_1 + \sigma_2( \boldsymbol{u}_0 )\, \delta u_2
+  \sum_{1\le i,j\le 3} \frac {\sigma_{i,j}(\boldsymbol{u}_0 )}{2} \, \delta u_i \delta u_j + \cdots =0 \, .
\end{equation}
For a generic $\boldsymbol{u}_0 \in W^{(1)}$ and $u_3$-flow ($\delta u_2=0$) this implies
\begin{equation} \label{exp1}
  \delta u_1 = \varkappa \, (\delta u_3)^2 + O\left( (\delta u_3)^3\right ), \qquad
\varkappa = -\frac{ \sigma_{33}(\boldsymbol{ u}_0)}{2 \sigma_1(\boldsymbol {u}_0 )} .
\end{equation}
The above expansion does not hold for the points on $W^{(1)}=\mathcal{C}$ with $\sigma_1 (\boldsymbol{u})=0$, i.e.,
at the points of $\mathcal{Z}\cap W^{(1)}$.

\begin{prop} \label{Z_W}
The subvariety $\mathcal{Z} =\{ \sigma(\boldsymbol{u})=0,\, \sigma_1(\boldsymbol{u})=0\}$ has precisely 3 common points with
$W^{(1)}$: $\mathcal{Z}\cap W^{(1)}=\{ \boldsymbol{r}_1, \boldsymbol{r}_2, \boldsymbol{r}_3 \}$.
\end{prop}

\noindent{\it Proof.} In view of \eqref{x_y_1}, along the stratum $W^{(1)}$ one has $\sigma_1(\boldsymbol{u})=0$
if and only if $x_1=0$,
which corresponds to the points $\boldsymbol{r}_1, \boldsymbol{r}_2, \boldsymbol{r}_3$. $\square$
\medskip

Next, it follows from Proposition \ref{sigmaexp}, that near $\boldsymbol{u} =\boldsymbol{r}_i$,
for the $u_3$-flow ($\delta u_2=0$), one has
\begin{equation} \label{exp_u1_u3}
\sigma (\boldsymbol{r}_i + {\bf \delta u}) = \sigma_2^{(i)} \left( -\frac{1}{\rho\sp{i}} (\delta u_1)^2
- \rho^i \delta u_1 \, \delta u_3 + \frac 12 (\delta u_3)^2 \right)+ O\left( \delta u_1 (\delta u_3)^2 \right)+
O\left((\delta u_3)^3 \right) =0,
\end{equation}
where, as above, $\rho = \exp(2\pi \imath/3)$.
Then the increment $\delta u_1$, as a function of $\delta u_3$, has 2 local branches
\begin{equation} \label{exp_2}
\delta u_1 = \frac 12 \rho^{2i}\left(-1 \pm \sqrt{3} \right) \delta u_3 + O \left( (\delta u_3)^2 \right) .
\end{equation}
Indeed, substituting this into the expansion \eqref{exp_u1_u3}, up to cubic terms, we get
$$
 \sigma (\boldsymbol{r}_i + {\bf \delta u}) = \pm \frac{\sigma_2^{(i)}}{2} (\rho^{3i}-1) (\delta u_3)^2 ,
$$
which is zero for any $i=1,2,3$.

Note also that $\boldsymbol{r}_1+\boldsymbol{r}_2+ \boldsymbol{r}_3\equiv 0$ in $\Jac(\mathcal{C})$
being the image of the divisor of $x$ under the Abel map. Due to expansion \eqref{expan_difs}
of the differentials $\Omega_j$ near $\infty \in {\mathcal C}$, a tangent vector to $ W^{(1)}\subset W^{(2)}$
at the origin $u=0$, $T_0  W^{(1)}$, is $(0,0,1)^T$. Hence
the $u_3$-flow is tangent to $W^{(1)}$ at the origin. Next, in view of \eqref{exp_Om_rho},
$T_{\boldsymbol{r}_i} W^{(1)}=(\rho^{i},0, \rho^{2i})^T$, which means that the flow
is also tangent to the projection of $W^{(1)}$ onto the $(u_2, u_3)$-plane at $\boldsymbol{r}_i$.
All these observations are depicted in Figure 5.1.
\begin{figure}[h,t]
\begin{center}
\includegraphics{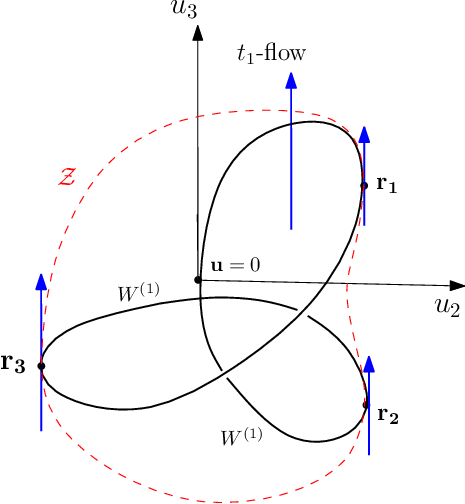}
\end{center} \caption{A sketch of the stratum $W^{(1)}\subset W^{(2)}$ (solid line) and $\mathcal Z$
(dashed line) in the projection onto the $(u_2, u_3)$-plane, and the $u_3$-flow on $W^{(2)}$.}
\end{figure}

\subsection{Expansions of the solutions near $W^{(1)}$ along the $u_3$-flow.}
The order of poles of the variables $J_i, \gamma_i$, as functions of $t_1$ (or $u_3$) depend on
the nature of the intersection (for example, transversal or tangential) of the $u_3$-flow with $W^{(1)}\subset W^{(2)}$.
The expansions of the solutions \eqref{sols_x} in powers of $u_3$ may be found by using
the corresponding expansions of the sigma-function and its derivatives near a point
$\boldsymbol{u}_0 \in W^{(1)}$, as well as the expansions \eqref{exp1}, \eqref{exp_2}.
It is more convenient however to find the expansions of the coordinates $(x_1,y_1), (x_2, y_2)$ of the points
$P_1, P_2\in S$ under the incomplete Abel map \eqref{quads2} and then use the formulae \eqref{solutioins_x_y}.

To do this, first note that, according to the definition of $W^{(1)}$,
when $\bf u$ belongs to $W^{(1)}\setminus \{ 0 \}$,
one of the points $P_i$ on $\mathcal{C}$, say $P_1$, coincides with $\infty$, whereas $P_2$ remains finite.
Now let $\Gamma(t)$, $|t-t_0| <1$ be a complex analytic arc in $W^{(2)}$ such that
$\Gamma (t_0)\cap W^{(1)}=\boldsymbol{u}_0$,
and the projection of the arc onto ${\mathbb C}^2=(u_2,u_3)$ is a segment of a straight line $u_2=$const.
Let $D(t)=\{P_1(t), P_2(t) \}$ be a
divisor on $\mathcal{C}$ such that its Abel image gives $\Gamma(t)$.
Then $P_1(t_0)=\infty, P_2(t_0)=(x_0,y_0)$. 

\begin{thm} \label{exp_x_y} For a generic $\boldsymbol{u}_0 =(u_{10},u_{20}, u_{30})\in W^{(1)}$,
near $t=t_0$ the coordinates of the points $P_1, P_2$ admit the expansion
\begin{equation}\label{exp_11}
\begin{aligned}
x_1 & = \frac {1}{(u_3-u_{30})^3} - \frac{3 y_0}{2x_0} \frac {1}{(u_3-u_{30})^2}
+ O\left((u_3-u_{30})^{-1}\right)  \\
 &= \frac{27}{8}\imath \frac{1}  {(t-t_0)^3} + \frac{27}{8}\frac{y_0}{x_0} \frac{1}{(t-t_0)^{2}}+
O\left( (t-t_{0})^{-1}\right) , \\
y_1 & =  \frac {1}{(u_3-u_{30})^4} + \frac{2 y_0}{x_0}\frac {1}{(u_3-u_{30})^3} + O((u_3-u_{30})^{-2}) \\
 & =  \frac {81}{16 (t-t_0)^4} + \frac{27\imath y_0}{4 x_0 }\frac {81}{16 (t-t_0)^3}  + O((t-t_0)^{-2}),  \\
x_2 & =  x_{0} - \frac{1}{2 \phi_{20}} (u_3-u_{30})^2 + O((u_3-u_{30})^3) ,  \\
 y_2 & = y_{0} +  O((u_3-u_{30} )^{2}).
\end{aligned}
\end{equation}
Next, near each point $\boldsymbol{u}_0 =\boldsymbol{r}_i$, $i=1,2,3$, at which $x_{20}=0$, there are
two expansions:
\begin{equation}
\begin{aligned}
x_1 & = \frac{1}{g^3} \frac {1}{ (u_3-u_{30})^3 } -\frac{3\rho^i}{2 g^3}\frac {1}{(u_3-u_{30})^2}+ O((u_3-u_{30})^{-1}),  \\
y_1 & = \frac{1}{g^4} \frac {1}{ (u_3-u_{30})^4 } -\frac{2\rho^i}{g^4}\frac {1}{(u_3-u_{30})^3}+ O((u_3-u_{30})^{-2}),  \\
x_2 & = \dfrac {3}{2} (-1\pm \sqrt{3})\rho^i \,(u_3-u_{30}) - \frac{3}{\rho^i} g \,(u_3-u_{30})^2 + O((u_3-u_{30})^{3}), \\
y_2 & = \rho^i + \frac{3}{4} (\mu_6-\mu_3\rho^i)\,(-1 \pm \sqrt{3})^2\, (u_3-u_{30})^2 + O((u_3-u_{30})^{3}),
\end{aligned}
\label{exp222}
\end{equation}
where $g=(3\mp \sqrt{3})/2$.
\end{thm}

The proof is given in Appendix.

Theorem \ref{exp_x_y} implies that when the $u_3$-flow crosses the substratum $W^{(1)}$ at a
generic point $\boldsymbol{u}_0$, we have
\begin{gather}
\frac{x_2y_1-x_1y_2}{x_2-x_1} = \frac{x_0}{ u_3-u_{30} }+ O(1), \quad
\frac{y_1-y_2}{x_2-x_1}= -\frac{1}{u_3-u_{30}} + O(u_3-u_{30}), \notag \\
x_1+x_2=\frac{1}{ (u_3-u_{30})^3 }+ O((u_3-u_{30})^{-2}) ,
\quad x_1 x_2= \frac{x_0}{(u_3-u_{30})^3}+ O((u_3-u_{30})^{-2}).  \label{elem_exp}
\end{gather}
In the case of crossing $W^{(1)}$ at the points $\boldsymbol{r}_i\in W^{(1)}$, one has instead
\begin{gather}
\begin{aligned}
\frac{x_2y_1-x_1y_2}{x_2-x_1} & = O(1)+ O((u_3-u_{30})) , \quad
\frac{y_1-y_2}{x_2-x_1}= -\frac{1} {g (u_3-u_{30})}+ O(1) , \\
x_1+x_2 &= \frac{1}{g^3 (u_3-u_{30})^3} + O(u_3-u_{30}) , \quad
x_1 x_2= \frac{3\rho^{i} (-1\pm \sqrt{3})}{2 g^3(u_3-u_{30})^2 }+ O(1).
\end{aligned} \label{elem_exp2}
\end{gather}
Note that the orders of the above expansions are compatible with those predictable from
the sigma-function solutions \eqref{sigma_quo}, \eqref{symm1}, \eqref{symm2}.

We stress that in all the cases the above symmetric functions of $x_i, y_i$, as functions of $u_3$ or $t_1$,
do not have poles  with branching.
However, they have finite branching along the intersection with $\mathcal{Z}\subset W^{(1)}$.

Using \eqref{sols_x} and the above expansions, one can estimate the leading terms of
formal series solutions for the variables $J_1\pm \imath J_2, \gamma_2\pm \imath \gamma_1, J_3, \gamma_3$ near the poles.
In the generic case one has
\begin{gather}
\gamma_3 = O\left( 1/(t-t_0)^{3/2} \right), \quad J_3= O\left( 1/(t-t_0) \right), \notag \\
\gamma_2+ \imath \gamma_1 = O(1),  \quad \gamma_2-\imath \gamma_1 = O\left( 1/(t-t_0)^{3} \right) , \label{principal_b} \\
 J_1 + \imath J_2 = O\left( 1/(t-t_0)^{5/2} \right), \quad  J_1 -\imath J_2 = O\left((t-t_0)^{1/2} \right), \notag
\end{gather}
and when the $u_3$-flow crosses the substratum $W^{(1)}$ at $\boldsymbol{r}_i$,
\begin{gather}
\gamma_3 = O( 1 ), \quad J_3= O\left( 1/(t-t_0) \right), \notag \\
\gamma_2+ \imath \gamma_1 = O(t-t_0),  \quad \gamma_2-\imath \gamma_1 = O\left( 1/(t-t_0) \right) ,
\label{secondary_b} \\
 J_1 + \imath J_2 = O\left( 1/(t-t_0)\right), \quad  J_1 -\imath J_2 = O(t-t_0). \notag
\end{gather}

These expansions correspond precisely with the leading behavior of formal series solutions of the Goryachev system
obtained directly by Kovalevskaya--Painlev\'e analysis. To observe this, first
rewrite the Goryachev system  \eqref{eq_Gor} with the Hamiltonian $H_1$ in the form
\begin{equation}
\begin{aligned}
\frac{d}{d \,t_1} (\gamma_2- \imath \gamma_1)
& = - 2\imath \frac 43 J_3 (\gamma_2- \imath \gamma_1) + 2\gamma_3 (J_1+ \imath J_2), \\
\frac{d}{d \,t_1} (\gamma_2+ \imath \gamma_1)
& = \; 2\imath \frac 43 J_3 (\gamma_2 + \imath \gamma_1) + 2\gamma_3 (J_1- \imath J_2), \\
 \frac{d}{d \,t_1} (J_1 + \imath J_2) & = -\frac 23 J_3 \imath (J_1 + \imath J_2)
-\frac 23 \frac{a\gamma_1+b}{ \gamma_3^{5/3} } (\gamma_2-\imath \gamma_1)+ \imath \,a \gamma_3^{1/3}, \\
 \frac{d}{d \,t_1} (J_1 - \imath J_2) & = \; \frac 23 J_3 \imath (J_1 - \imath J_2)
-\frac 23 \frac{a\gamma_1+b}{ \gamma_3^{5/3} } (\gamma_2 + \imath \gamma_1) - \imath\, a \gamma_3^{1/3}, \\
 \frac{d}{d \,t_1}  J_3 & = - a \frac {\gamma_2}{\gamma_3^{2/3}},  \\
 \frac{d}{d \,t_1}\, \gamma_3 & = -(J_1+ \imath J_2)(\gamma_2+ \imath \gamma_1) - (J_1-\imath J_2)(\gamma_2-\imath \gamma_1).
\end{aligned} \label{eqn_new_form}
\end{equation}
Then the corresponding formal Puiseaux (or Laurent) series solutions of \eqref{eqn_new_form} are of 2 kinds.
According to terminology of \cite{AvM1}, those depending on the maximal number of free parameters (here three,
for example, the constants of motion $h_1$, $h_2$ and a local coordinate on $W^{(1)}$)
represent the {\it principle balances} of the solutions. Their leading behavior coincides with \eqref{principal_b}.
The series solutions depending on 2 or less free parameters (called {\it secondary balances}) correspond to
\eqref{secondary_b}.

\section*{Appendix.}
\subsection*{Proof of Proposition \ref{sigmaexp}.}
Let, as above, $q_i=(0,\rho^i)\in \mathcal{C}$,
$\boldsymbol{r}_i=\int_\infty^{q_i} \boldsymbol{\Omega}$.
Since $\boldsymbol{r}_{i}\in W^{(1)}$ and $x(q_i)=0$, Proposition \ref{sigmas_on_W} implies that
$\sigma_{1}( \boldsymbol{r}_{i})=\sigma_{3} ( \boldsymbol{r}_{i} )=0$.
Let $\xi$ be the local coordinate on $\mathcal{C}$ near $\infty$
and $x$ be such coordinate near $q_i=(0,\rho^i)\in \mathcal{C}$. Introduce the functions
$$
\boldsymbol{U}(\xi)=(U_1, U_2, U_3)= \int_\infty^P \boldsymbol{\Omega}, \quad \text{and} \quad
\boldsymbol{w}(x)=(w_1,w_2, w_3)= \int_{q_i}^P \boldsymbol{\Omega},
$$
In view of the expansions \eqref{expan_difs}, we obtain
\begin{align}
U_1 & = - \frac 15 \xi^5 -\frac{1}{21}\mu_2 \xi^7 + \cdots, \notag \\
U_2 & = - \frac 12 \xi^2 -\frac{1}{12}\mu_2 \xi^4 + \frac{2}{15} \mu_3 \xi^5+ \cdots, \\
U_3 & = -\xi + \frac{1}{12}\mu_3 \xi^4 + \frac{1}{45}\mu_2^2 \xi^5 + \cdots,  \notag
\end{align}
and, in view of \eqref{exp_Om_rho}, for $i=1$ we have
\begin{align*}
w_1 & ={\displaystyle \frac {\rho }{3}} \,x - {\displaystyle \frac {1}{27}}
\,{\displaystyle \frac {2\,\rho ^{2}\,{\mu _{6}} - {\mu_2}}{\rho }} \,x^{3} - {\displaystyle \frac {1}{18}} \,\rho \,{\mu
_{3}}\,x^{4} + {\displaystyle \frac {1}{135}} \,{\displaystyle
\frac {(3\,\rho ^{2}\,{\mu _{6}} - 2\,{\mu _{2}})\,{\mu _{6}}}{
\rho }} \,x^{5} + \mathrm{O}(x^{6}), \\
w_2 & = {\displaystyle \frac {\rho }{6}} \,x^{2} - {\displaystyle \frac {
1}{36}} \,{\displaystyle \frac {2\,\rho ^{2}\,{\mu _{6}} - {\mu
_{2}}}{\rho }} \,x^{4} - {\displaystyle \frac {2}{45}} \,\rho \,{
\mu _{3}}\,x^{5} + \mathrm{O}(x^{6}), \\
w_3 & ={\displaystyle \frac {\rho ^{2}}{3}} \,x - {\displaystyle \frac {
1}{27}} \,\rho ^{2}\,{\mu _{6}}\,x^{3} - {\displaystyle \frac {1
}{36}} \,{\mu _{3}}\,\rho ^{2}\,x^{4} + {\displaystyle \frac {1}{
135}} \,{\displaystyle \frac {\rho ^{4}\,{\mu _{6}}^{2} + \rho ^{
2}\,{\mu _{6}}\,{\mu _{2}} - {\mu _{2}}^{2}}{\rho ^{2}}} \,x^{5}
 + \mathrm{O}(x^{6}) .
\end{align*}
The expansions for $i=2,3$ are obtained from the above by replacing $\rho$ by
$\rho^2$ and $\rho^3=1$ respectively.

Now note that for any $x,\xi\in {\mathbb C}$ we have $\boldsymbol{r}_i + \boldsymbol{w} (x)\in W^{(1)}$ and
$\boldsymbol{r}_i + \boldsymbol{w} (x) + \boldsymbol{U} (\xi) \in W^{(2)}$.
Hence, in view of Proposition \ref{sigmas_on_W},
$$
\Sigma (x,\xi) := \sigma (\boldsymbol{r}_i + \boldsymbol{w} (x) + \boldsymbol{U} (\xi)  ) \equiv 0, \quad
\Sigma_3 (x)=\sigma_3 (\boldsymbol{r}_i + \boldsymbol{w} (x)) \equiv 0\, .
$$
Substituting the above two expansions into the sigma-expansions \eqref{sigma_r} for $i=1$, we obtain\footnote{
Here we omit the index $(i)$ in the $\sigma$-derivatives.}
\begin{gather*}
\Sigma(x,\xi) = \left({\displaystyle \frac {1}{6}} \,{\sigma _{2}}\,
\rho  + {\displaystyle \frac {1}{18}} \,{\sigma _{3, \,3}}\,\rho^{4}
+ {\displaystyle \frac {1}{18}} \,{\sigma _{1, \,1}}\,\rho^{2} +
{\displaystyle \frac {1}{9}} \,{\sigma _{1, \,3}}\,\rho^{3} \right)\,x^{2} \\
- \frac {1}{3} (\sigma_{3, \,3}\,\rho ^{2} + \sigma_{1, \,3}\,\rho )\,\xi \,x
 + ( - {\displaystyle \frac {1}{2}} \,\sigma_{2} + {\displaystyle \frac {1}{2}} \,{\sigma _{3, \,3}})\,\xi ^{2}
 \\
+ \left( {\displaystyle \frac {1}{54}} \,{\sigma _{1, \,3, \,3}}\,\rho ^{5
} + {\displaystyle \frac {1}{18}} \,{\sigma _{1, \,2}}\,\rho ^{2}
 + {\displaystyle \frac {1}{162}} \,{\sigma _{3, \,3, \,3}}\,\rho
 ^{6} + {\displaystyle \frac {1}{54}} \,{\sigma _{1, \,1, \,3}}\,
\rho ^{4} + {\displaystyle \frac {1}{18}} \,{\sigma _{2, \,3}}\,
\rho ^{3} + {\displaystyle \frac {1}{162}} \,{\sigma _{1, \,1, \,
1}}\,\rho ^{3}\right )x^{3} \\
 - \left( \frac {1}{18} \,{\sigma_{1,\,1, \,3}}\,\rho ^{2}
+ \frac {1}{9} \,\sigma_{1, \,3, \,3}\,\rho ^{3} + \frac {1}{6} \,\sigma_{2, \,3}\,\rho +
\frac {1}{18} \,{\sigma _{3, \,3, \,3}}\,\rho ^{4} \right) \,\xi \,x^{2} \\
 + \left( - \frac 1 6 \,\sigma_{2, \,3}\,\rho ^{2} + \frac {1}{6} \,\sigma _{1, \,3, \,3}
\,\rho  - \frac {1}{6} \,{\sigma_{1, \,2}}\,\rho + \frac {1}{6} \,{\sigma _{3, \,3, \,3}}\,
\rho ^{2} \right)\,\xi ^{2}\,x + \left( \frac {1}{2} \,\sigma_{2, \,3}
- \frac {1}{6} \,\sigma_{3, \,3, \,3}\right)\,\xi^{3} + O(\xi^3 x )
\end{gather*}
and
$$
\Sigma_3 (x)= \left(\frac {1}{3} \,\sigma_{1, \,3}\,\rho  +
\frac {1}{3} \,\sigma_{3, \,3}\,\rho^{2} \right) \,x
 + \left( \frac {1}{18} \, \sigma_{3, \,3, \,3}\,\rho^{4} +
\frac {1}{18} \,{\sigma _{1, \,1, \,3}}\,\rho ^{2} +
\frac {1}{9} \,{\sigma _{1, \,3, \,3} }\,\rho ^{3} + \frac {1}{6} \,{\sigma _{2, \,3}}
\,\rho \right )\,x^{2} + O(x^3).
$$
Then, upon equating the coefficients of the expansions $\Sigma(x,\xi), \Sigma_3(x)$ to zero, we get a system of equations for the
coefficients $\sigma_{i,\dots,k}^{(\alpha)}$. (Note that
the indicated coefficients of $\Sigma_3(x)$ are also coefficients of $\Sigma(x,\xi)$,
so they, in fact, do no bring new conditions.) Solving it, we obtain the first group of relations \eqref{s_sr}.

To find the relations \eqref{ders_rho}, we proceed in the same manner as in \eqref{Hopital1} and consider the limit
$$
Q_i = \lim\limits_{ \bf{u} \to 0}
\frac{ \sigma_1( \boldsymbol{r}_i + {\bf u})
+ \rho^i \sigma_i( \boldsymbol{r}_i+ {\bf u})}{ \sigma ( \boldsymbol{r}_i+ \bf{u}) }.
$$
Applying l'Hopital's rule, we get
\begin{align*}
Q_i & = \lim\limits_{u_3 \to 0}
\frac{ \sigma_1 (\boldsymbol{r}_i+ u_3 {\bf e_3} ) + \rho^i\, \sigma_i ( \boldsymbol{r}_i+ u_3{\bf e_3} )}
{\sigma ( \boldsymbol{r}_i+ u_3 {\bf e_3} )}=
\lim\limits_{u_3 \to 0}
\frac{ \sigma_{13}(\boldsymbol{r}_i+ u_3{\bf e_3} )+
\rho^i\,  \sigma_{33}(\boldsymbol{r}_i+ u_3{\bf e_3} )}{\sigma_3( \boldsymbol{r}_i+ u_3{\bf e_3} ) } \\
& =\frac{\sigma_{133}(\boldsymbol{r}_i)+ \rho^i\, \sigma_{333}(\boldsymbol{r}_i) }
{\sigma_{33}(\boldsymbol{r}_i)}, \qquad {\bf e_3} =(0,0,1)^T.
\end{align*}
On the other hand,
\begin{align*}
Q_i & = \lim\limits_{u_2 \to 0}
\frac{ \sigma_1(\boldsymbol{r}_i+ u_2 {\bf e_2} )+ \rho^i \,
\sigma_3 ( \boldsymbol{r}_i+ u_2{\bf e_2} )}{\sigma ( \boldsymbol{r}_i+ u_2 {\bf e_2})}=
\frac{ \sigma_{12}(\boldsymbol{r}_i)+ \rho^i \,
\sigma_{23}(\boldsymbol{r}_i)}{\sigma_2(\boldsymbol{r}_i) } , \qquad {\bf e_2}=(0,1,0)^T,  \\
Q_i & = \lim\limits_{u_1 \to 0}
\frac{ \sigma_1(\boldsymbol{r}_i+ u_1{\bf e_1})
+ \rho^i \,\sigma_3 ( \boldsymbol{r}_i+ u_1 {\bf e_1})}{\sigma ( \boldsymbol{r}_i+ u_1 {\bf e_1}) }=
\lim\limits_{u_1 \to 0}
\frac{ \sigma_{11}(\boldsymbol{r}_i+ u_1{\bf e_1} )
+ \rho^i \,\sigma_{13}(\boldsymbol{r}_i+ u_1 {\bf e_1})}{\sigma_1(\boldsymbol{r}_i+ u_1 {\bf e_1} ) } \\
& =\frac{ \sigma_{111}(\boldsymbol{r}_i)+ \rho^i\, \sigma_{113}(\boldsymbol{r}_i)}{\sigma_{11}(\boldsymbol{r}_i)},
\qquad {\bf e_1}=(1,0,0)^T.
\end{align*}
Comparing the above limits and using \eqref{s_sr}, we get
\begin{gather*}
\sigma_{133}(\boldsymbol{r}_3)+ 3 \rho^{i} \sigma_{23} (\boldsymbol{r}_3)
= \sigma_{12}(\boldsymbol{r}_3)+ \rho^{i} \sigma_{23}(\boldsymbol{r}_3), \\
-\rho^{i} \frac{\sigma_{111}(\boldsymbol{r}_3) + \rho^{i}\sigma_{113}(\boldsymbol{r}_3) }
{2 \sigma_{2}(\boldsymbol{r}_3) }
= \frac{\sigma_{12}(\boldsymbol{r}_3) + \rho^{i} \sigma_{23}(\boldsymbol{r}_3) }{\sigma_{2}(\boldsymbol{r}_3) } .
\end{gather*}
These relations are compatible with a vanishing of the leading
coefficients of the expansions $\Sigma (x,\xi), \Sigma_3 (x)$ if and only if relations \eqref{ders_rho} hold.
This proves the proposition.
\medskip

\noindent{\it Proof of Theorem} \ref{exp_x_y}.
It is sufficient to study the expansion of the map \eqref{quads2} near the point
$\{\infty, (x_0,y_0)\}\in \mathcal{C}\times \mathcal{C}$.
 Again let $\xi=x^{-1/3}$ be a local coordinate of $P_1\in \mathcal{C} $ near $\infty$ and $\kappa=x-x_0$ be such a coordinate
of $P_2$ near $(x_0,y_0)$, $x_0\ne 0$.
The expansions of the differentials $\Omega_i$ near $(x_0, y_0)$ are
\begin{align*}
\Omega_2 & = \frac{x}{\partial G/\partial y(x,y)}dx
      =\left(\varphi_{20} + \varphi_{21}\kappa+ O(\kappa^2) \right) d\kappa, \\
\Omega_3 & = \frac{y}{\partial G/\partial y(x,y)}dx
=\left( \varphi_{30} + \varphi_{31}\kappa+ O(\kappa^2)\right)  d\kappa,
\end{align*}
where $\varphi_{20},\dots, \varphi_{31}$ are certain nonzero constants depending on $x_0$ and the coefficients of the curve.
Note that $\frac{\phi_{30}}{\phi_{20}}=\frac{y_0}{x_0}$.
In view of this and of the expansions of $\Omega_i$ in \eqref{expan_difs}, the differential of the map
\eqref{quads2} reads
\begin{equation}
\label{tau_x0}
\begin{pmatrix} \xi + \dfrac{\mu_2}{3}\xi^3 + O(\xi^4) & \varphi_{20} + \varphi_{21}\kappa+ O(\kappa^2)  \\
                1 -\dfrac{\mu_3}{3} \xi^3 + O(\xi^4) & \varphi_{30} + \varphi_{31}\kappa+ O(\kappa^2) \end{pmatrix}
\begin{pmatrix} d \xi \\ d \kappa \end{pmatrix} = \begin{pmatrix} du_2 \\ du_3 \end{pmatrix}\, .
\end{equation}
Taking into account that $\varphi_{20}, \varphi_{30}$ are non-zero, we may invert the above matrix expansion yielding
(up to quadratic terms)
\begin{equation} \label{inv_dif}
\begin{pmatrix} d \xi \\ d \kappa \end{pmatrix} =
\begin{pmatrix} -\dfrac{\phi_{30} } {\phi_{20}} - \dfrac{\phi_{30}^2} {\phi_{20}^2} \xi- \Phi \kappa &
1+ \dfrac{\phi_{30} } {\phi_{20}} \xi \\ 
  \dfrac{1}{\phi_{20}} + \dfrac{\phi_{30}}{\phi_{20}^2}\xi -\dfrac{\phi_{21}}{\phi_{20}^2}\kappa 
&  -\dfrac{1}{ \phi_{20} } \xi
 \end{pmatrix}
\begin{pmatrix} du_2 \\ du_3 \end{pmatrix}\, , \qquad
\Phi= \frac{\phi_{31}\phi_{20}-\phi_{21}\phi_{30}}{\phi_{20}^2}.
\end{equation}

For the $t_1$-flow we have $du_3= \frac 23 \imath \,dt$ and $du_2=0$.
Then \eqref{inv_dif} gives a system of 2 ODEs with respect to $u_3-u_{30}$ (or $t-t_0$).
Applying the condition $\xi(0)=0, \kappa(0)=0$, we find initial terms of the series solutions
\begin{align*}
\xi & = u_3-u_{30} + \frac 12 \frac{y_0}{x_0} (u_3-u_{30})^2 + O((u_3-u_{30})^3)
= \frac 23 \imath (t-t_0)- \frac 12 \frac{y_0}{x_0} \frac {2^2}{3^2} (t-t_0)^2+ \cdots , \\
\kappa & = \quad - \frac{1}{2\phi_{20}} (u_3-u_{30})^2 + O((u_3-u_{30})^3)
= \frac{1}{2\phi_{20}}\frac {2^2}{3^2} (t-t_0)^2+ \cdots .
\end{align*}
Substituting this into the expansion \eqref{exp_y} and taking into account \eqref{mu's}, we get
the expansions \eqref{exp_11}. 

The above argument fails to work only when $P_2(t_0)= q_i=(0,\rho^i)$ and $\phi_{20}(q_i)=0$,
i.e., when the differential relation \eqref{tau_x0} cannot be locally inverted.
In this case we will use the expansions of the differentials $\Omega_i$ near $q_i$ given by \eqref{exp_Om_rho}
and replace \eqref{tau_x0} by the differential of the Abel map \eqref{AM2} taking the part with
$\Omega_1, \Omega_3$ and $u_1, u_3$. Let $x$ be a local coordinate on $\mathcal{C}$ near $q_i=(0,\rho^i)$.
For $\alpha=1$ the expansion of the differential is
\begin{equation}
\begin{pmatrix} \xi^4 + \dfrac{\mu_2}{3} \xi^6+  O(\xi^7) &
\dfrac {\rho}{3} - \dfrac {1}{9 \rho} (2 \rho^{2} \mu_{6}-\mu_2) x^2 +O(x^3) \\
1 -\dfrac{\mu_3}{3} \xi^3 + O(\xi^4) & \dfrac {\rho^2}{3} -\dfrac {\mu_6 \rho^2}{9}\,x^{2}+O(x^3)  \end{pmatrix}
\begin{pmatrix} d \xi \\ d x \end{pmatrix} = \begin{pmatrix} du_1 \\ du_3 \end{pmatrix} .
\end{equation}
The latter is invertible and, up to cubic terms in $\xi, x$, gives the expansions
\begin{equation}
\begin{pmatrix} d \xi \\ d x \end{pmatrix} =
\begin{pmatrix}
  - \rho - \rho^{2}\,\xi  + \frac {1}{3 \rho^2} \,
( - \mu _{6} + \rho \,\mu_2)\,x^{2} - \xi^{2} &
1 + \xi \,\rho  + \rho ^{2}\,\xi ^{2} \\
\frac {3}{\rho } + 3\,\xi  + (2\,{\mu _{6}}\,\rho ^{2}- {\mu _{2}})\,x^{2} + 3\,\rho \,\xi ^{2}
 &  - \dfrac {3\,\xi }{\rho } - 3\,\xi ^{2}
\end{pmatrix}
\begin{pmatrix} du_1 \\ du_3 \end{pmatrix}\, .
\end{equation}
According to \eqref{exp_2}, for $i=1$ one has $du_1 = \nu d u_3$, $\nu= \frac 12(-1 \pm \sqrt{3})\rho^2$.
Then the above expansions lead to the following system of 2 ODEs with respect to $u_3-u_{30}$
\begin{align*}
\frac{d \xi}{du_3} & = g(1+\rho \,\xi + \rho^2\,\xi^2) + \frac{\nu}{3\rho^2}(\rho \mu_2-\mu_6)x^2+O(\xi^3) \\
\frac{d x}{du_3} & = \frac{3\nu}{\rho} - \frac{3 g}{\rho} \xi - 3 g \xi^2+ (2 \rho^2 \mu_6-\mu_2) \nu \, x^2
+ O(\xi x^2),
\end{align*}
where
$$
g=1-\rho \nu=1-\frac 12(-1 \pm \sqrt{3}) =\frac {3 \mp \sqrt{3}}{2}.
$$
Applying again the conditions $\xi(0)=0, x(0)=0$, we find the series solutions with initial terms
\begin{align*}
\xi & =g (u_3- u_{30}) + \frac{\rho\,g}{2} (u_3- u_{30})^2 + O\left((u_3- u_{30})^3 \right),  \\
x & =\frac{3\nu}{\rho} (u_{3}-u_{30}) - \frac{3g}{\rho} (u_{3}-u_{30})^2 +O\left( (u_{3}-u_{30})^3 \right) ,
\end{align*}
which, in view of the expansions \eqref{exp_y}, \eqref{y_r}, give \eqref{exp222}. The expansions for $i=2,3$
are obtained by replacing $\rho$ by $\rho^2$ and by $\rho^3=1$ respectively. $\square$

\section*{Acknowledgments}
The work of Yu.F was supported by the MICIIN grants MTM2009-06973 and MTM2009-06234.
Each author is grateful to ZARM, Bremen University, and the Department of Physics,
Oldenburg University, for funding research visits to these institutions and enabling
the meetings of the authors that allowed the completion of the present article.
We also thank S. Abenda and A. Tsiganov for stimulating discussions.

\end{document}